\documentclass[journal,10pt]{IEEEtran}

\usepackage{stfloats}
\usepackage[cmex10]{amsmath}
\usepackage{amssymb,latexsym}
\usepackage{graphicx}
\usepackage{color,cite,times}
\usepackage{epstopdf}
\usepackage{algorithmic}
\usepackage[ruled,linesnumbered] {algorithm2e}
\usepackage{booktabs}
\usepackage{url}
\usepackage{fancyhdr}
\usepackage{cleveref}
\usepackage{verbatim}
\usepackage{multirow}
\usepackage{makecell}
\usepackage{graphicx}
\usepackage{indentfirst}
\usepackage{cases}
\usepackage{amsthm,amsmath}
\usepackage{extarrows}
\usepackage{threeparttable}
\usepackage{diagbox}
\usepackage{pifont}
\usepackage{subfigure}
\usepackage{bbm}
\usepackage{ragged2e}

\newcommand{\bsm}{\boldsymbol}

\newcommand{\mbf}{\mathbf}
\newcommand{\diag}{\mathrm{diag}}

\newtheorem{prop}{Proposition}

\newtheorem{remk}{Remark}
\newtheorem{lemm}{Lemma}

\begin{document}
\title{RIS Partitioning Based Scalable Beamforming Design for Large-Scale MIMO: Asymptotic Analysis and Optimization}
\author{Chang Cai, \IEEEmembership{Graduate Student Member, IEEE,}
Xiaojun Yuan, \IEEEmembership{Senior Member, IEEE,} \\
and Ying-Jun Angela Zhang, \IEEEmembership{Fellow, IEEE}
	\thanks{
		Manuscript received March 15, 2022; revised October 22, 2022; accepted January 19, 2023.
		This work was supported in part by the Sichuan Science and Technology Program under Grants 2022ZYD0120 and 2021YFH0014, and in part by the General Research Fund (project number 14201920, 14202421, 14214122) from the Research Grants Council of Hong Kong.
		This paper was presented in part at the IEEE Global Communications Conference (GLOBECOM) 2022, Rio de Janeiro, Brazil \cite{conference}.
		The associate editor coordinating the review of this article and approving it for publication was V. Sciancalepore.
		\textit{(Corresponding author: Xiaojun Yuan.)}
		
		Chang Cai and Ying-Jun Angela Zhang are with the Department of Information Engineering, The Chinese University of Hong Kong, Hong Kong (e-mail: cc021@ie.cuhk.edu.hk; yjzhang@ie.cuhk.edu.hk).
		
		Xiaojun Yuan is with the National Key Laboratory of Science and Technology on Communications, University of Electronic Science and Technology of China, Chengdu 611731, China (e-mail: xjyuan@uestc.edu.cn). 
}
}
\maketitle

\begin{abstract}
In next-generation wireless networks, reconfigurable intelligent surface (RIS)-assisted multiple-input multiple-output (MIMO) systems are foreseeable to support a large number of antennas at the transceiver as well as a large number of reflecting elements at the RIS.
To fully unleash the potential of RIS, the phase shifts of RIS elements should be carefully designed, resulting in a high-dimensional non-convex optimization problem that is hard to solve with affordable computational complexity.
In this paper, we address this scalability issue by partitioning RIS into sub-surfaces, so as to optimize the phase shifts in sub-surface levels to reduce complexity.
Specifically, each sub-surface employs a linear phase variation structure to anomalously reflect the incident signal to a desired direction, and the sizes of sub-surfaces can be adaptively adjusted according to channel conditions.
We formulate the achievable rate maximization problem by jointly optimizing the transmit covariance matrix and the RIS phase shifts.
Under the RIS partitioning framework, the RIS phase shifts optimization reduces to the manipulation of the sub-surface sizes, the phase gradients of sub-surfaces, as well as the common phase shifts of sub-surfaces.
Then, we characterize the asymptotic behavior of the system with an infinitely large number of transceiver antennas and RIS elements.
The asymptotic analysis provides useful insights on the understanding of the fundamental performance-complexity tradeoff in RIS partitioning design.
We show that in the asymptotic domain, the achievable rate maximization problem has a rather simple form with an explicit physical meaning of optimization variables.
We develop an efficient algorithm to find an approximately optimal solution to the asymptotic problem via a one-dimensional (1D)  grid search.
Moreover, we discuss the insights and impacts of the asymptotic result on finite-size system design.
By applying the asymptotic result to a finite-size system with necessary modifications, we show by numerical results that the proposed design achieves a favorable tradeoff between system performance and computational complexity.
\end{abstract}


\section{Introduction}
Emerging data-intensive applications, such as virtual reality, holographic projection, and autonomous driving, give rise to urgent needs for high-speed and seamless data services in future wireless systems \cite{letaief2019roadmap}.
A main bottleneck for the improvement of the quality of service (QoS) lies in the randomness and uncontrollability of wireless communication environments, in which the reliability of a link may severely deteriorate due to deep fading and shadowing effects.
Thanks to the recent advances on programmable metamaterials \cite{Nanfang2011Snell,cui2014coding}, reconfigurable intelligent surface (RIS) has emerged as a promising new technology to improve the link reliability of wireless networks as it can artificially configure the wireless channel in a favorable manner.
Typically, a RIS is a planar surface consisting of a large number of low-cost and passive reflecting elements.
By inducing an appropriately designed phase shift at each passive element, a RIS can manipulate the incident signals to be constructively or destructively superimposed at receiver (Rx) (referred to as \textit{passive beamforming}),
thereby reshaping the wireless channel to boost the performance of communication systems \cite{di2019smart, liaskos2018new, xjyuan2021magzine}.

Extensive research efforts \cite{QingqingWu2019TWC,HuayanGuo2020WSR,XianghaoYu2019ICCC,ShuowenZhang2020JSAC,CunhuaPan2020TWC,ChongwenHuang2019TWC,ycliang2021CR,ChangCai2020JCIN, yuanwei2020NOMA, hliu2021fl, zjzhang2021tsp} have been devoted to the optimization of passive beamforming based on various design criteria for RIS-aided communication systems.
Aiming to maximize the achievable rate, the joint optimization of transmit beamforming and RIS phase shifts (referred to as \textit{joint active and passive beamforming}) are considered for multiple-input single-output (MISO) systems in \cite{QingqingWu2019TWC,HuayanGuo2020WSR,XianghaoYu2019ICCC} and for multiple-input multiple-output (MIMO) systems in \cite{ShuowenZhang2020JSAC, CunhuaPan2020TWC}.
Ref. \cite{ChongwenHuang2019TWC} developed a RIS power consumption model to study the energy efficiency maximization problem, which shows that a passive RIS is more energy efficient than an amplify-and-forward (AF) relay.
Thanks to its ability to suppress interference, RIS is also widely exploited in cognitive radio networks \cite{ycliang2021CR}, device-to-device (D2D) communications \cite{ChangCai2020JCIN}, and non-orthogonal multiple access (NOMA) \cite{yuanwei2020NOMA}.
Moreover, the authors in \cite{hliu2021fl} demonstrated the effectiveness of passive beamforming on improving the over-the-air federated learning (FL) performance.

The passive beamforming design mentioned above, however, faces a serious scalability issue in real implementation.
On one hand, the transmitter-RIS-receiver (Tx-RIS-Rx) link suffers from the \textit{double fading effect}, i.e., the equivalent path loss of the Tx-RIS-Rx link is the product (instead of the sum) of the path losses of the Tx-RIS and RIS-Rx links \cite{WankaiTang2020TWC_PathLoss}.
The double fading effect results in severe path loss of the Tx-RIS-Rx link.
Thus, to achieve a substantial passive beamforming gain, a large RIS with hundreds and thousands of reflecting elements is needed.
On the other hand, due to the non-convex unit-modulus constraint imposed on RIS phase shifts, the computational complexity involved in passive beamforming optimization is at least cubic in the number of RIS elements \cite{QingqingWu2019TWC,ChongwenHuang2019TWC,HuayanGuo2020WSR,XianghaoYu2019ICCC,ShuowenZhang2020JSAC,CunhuaPan2020TWC,ycliang2021CR,ChangCai2020JCIN, yuanwei2020NOMA, hliu2021fl, zjzhang2021tsp}.
This implies a prohibitively high computational complexity even for a RIS with a typical size, which may seriously impede the widespread application of the RIS technology in next-generation wireless communications. 

To address the scalability issue, \cite{yfyang2020ofdm, bxzheng2020wcl, cjzhong2022elementgrouping, ccai2021wcl, physics2021tcom, basar2021partitioning} proposed to partition a RIS into sub-surfaces, where each sub-surface consists of a number of adjacent reflecting elements.
Specifically, the authors in \cite{yfyang2020ofdm, bxzheng2020wcl, cjzhong2022elementgrouping} assumed a common phase shift of elements in each sub-surface, thereby reducing the number of passive beamforming variables to that of the sub-surfaces.
More recently, the authors in \cite{ccai2021wcl, physics2021tcom} considered a linear phase variation structure in sub-surfaces, where a phase gradient and a common phase shift of each sub-surface can be adjusted for passive beamforming.
By employing the linear phase variation structure, a sub-surface can work as an anomalous reflector to change the direction and wavefront of the reflected beam.
Clearly, RIS partitioning exhibits a tradeoff between passive beamforming gain and computational complexity, which can be flexibly controlled by adjusting the number of partitioned sub-surfaces.
However, the existing approaches in \cite{yfyang2020ofdm, bxzheng2020wcl, cjzhong2022elementgrouping, physics2021tcom, ccai2021wcl, basar2021partitioning} are all based on heuristic optimization methods that provide limited insights into the utmost potential of the RIS partitioning technique.
As such, it is of pressing importance to develop an analytical framework to characterize the fundamental performance-complexity tradeoff of RIS partitioning, which motivates the work presented in this paper.

Meanwhile, the use of large antenna arrays at both the Tx and Rx ends, referred to as \textit{large-scale MIMO}, has recently attracted substantial research attention due to its great potential to improve spectral efficiency and spatial resolution \cite{largescale_MIMO_heath, largescale_MIMO_yuwei, largescale_MIMO_shijin, largescale_MIMO_yongzeng}.
To name a few, \cite{largescale_MIMO_heath} proposed a compressed-sensing-based algorithm to solve the hybrid beamforming problem in the point-to-point large-scale MIMO system.
The proposed algorithm leverages the sparse nature of mmWave channels and achieves a performance close to the fully digital beamforming baseline.
Ref. \cite{largescale_MIMO_shijin} characterized the ergodic sum rate for large-scale MIMO multiple access channels.
The derivations are based on large dimensional random matrix theory, assuming that the numbers of transceiver antennas approach infinity with a fixed ratio. 
Ref. \cite{largescale_MIMO_yongzeng} provided a unified modeling for large-scale MIMO channels by taking the near-field radiation and the physical size of array elements into account.
It is worth noting that the interplay between RIS and large-scale MIMO can undoubtedly further improve the performance of future wireless networks.
This interplay, however, brings additional challenges to system design since the active and passive beamforming should be jointly optimized in a scalable manner.

In this paper, we study joint active and passive beamforming to maximize the achievable rate of the RIS-aided large-scale MIMO system.
We take the RIS partitioning approach and provide a scalable solution to the problem of joint active and passive beamforming design.
In our RIS partitioning design, each sub-surface adopts a linear phase variation structure to achieve anomalous reflection, and the size of each sub-surface can be adaptively adjusted based on channel conditions.
The RIS optimization then reduces to the manipulation of the sub-surface sizes, the phase gradients of sub-surfaces, as well as the common phase shifts of sub-surfaces.
Different from the existing works \cite{yfyang2020ofdm, bxzheng2020wcl, cjzhong2022elementgrouping, physics2021tcom, ccai2021wcl, basar2021partitioning}, we focus on the asymptotic regime where the number of RIS reflecting elements and the number of transceiver antennas go to infinity.
In this regime, the RIS and the transceiver arrays have infinitely high resolution to distinguish the Tx-RIS and RIS-Rx channel paths with different arrival and departure angles.
We show that the Tx-RIS-Rx channel paths are asymptotically orthogonal to each other, implying that arbitrary common phase shifts of the sub-surfaces are optimal in the asymptotic regime.
Based on that, we establish the asymptotic form of the rate maximization problem involving the sub-surface sizes, the phase gradients of sub-surfaces, and the power allocation among different channel paths at the Tx side.

We establish the analytical solution to the above asymptotic rate maximization problem.
Specifically, by adopting the linear phase variation structure, each sub-surface needs to reflect the incident signals from a Tx-RIS arrival path to a RIS-Rx departure path, or in other words, the optimization of the sub-surface phase gradients reduces to a Tx-RIS-Rx path pairing problem.
We show that the optimal path pairing strategy can be expressed in closed form.
Furthermore, 
by analysing the Karush-Kuhn-Tucker (KKT) conditions of the problem, 
we show that the global optimum of the RIS partition sizes and the Tx power allocation is characterized by the solution to a set of non-linear scalar equations.
To solve the equations, we propose a one-dimensional (1D) grid search algorithm which outputs an approximately optimal solution.
As a low-complexity alternative, we outline the Levenberg-Marquardt (LM) method \cite{more1978levenberg} which reaches a stationary point.
Interestingly, the optimal RIS partitioning strategy allows for a water-filling-like interpretation.
That is, the RIS is dedicated to serving the strongest Tx-RIS and RIS-Rx path pair in the low signal-to-noise (SNR) regime, and is evenly partitioned to serve every paired Tx-RIS-Rx path in the high SNR regime.

We further discuss the insights and impacts of the above asymptotically optimal solution on finite-size system design.
We first discuss how to adapt the asymptotic solution to a finite-size system with practical antenna and RIS settings.
Then, we show that the proposed RIS partitioning approach only has a marginal performance loss compared to the conventional element-wise optimization method \cite{CunhuaPan2020TWC}, whereas the former reduces the computational time by orders of magnitudes especially when the number of RIS elements is large.
Therefore, our proposed RIS partitioning approach strikes an appealing balance between performance and complexity, and provides a scalable solution to the design of large-scale RIS-aided systems.

The rest of this paper is organized as follows.
Section \ref{section-system-model} describes the system model of the RIS-aided large-scale MIMO communication with RIS partitioning, and then formulates the achievable rate maximization problem.
Section \ref{sec-asymptotic-formulation} derives the asymptotic formulation of the problem.
Section \ref{sec-algorithm} elaborates the algorithm design to solve the asymptotic problem, and then discusses the method to construct a feasible solution for finite-size systems.
Section \ref{sec-simluation} provides the simulation results.
Finally, we conclude this paper in Section \ref{sec-conclusion}.

\textit{Notations:}
Lower-case letters are used to denote scalars.
Vectors and matrices are denoted by lower-case and upper-case boldface letters, respectively.
$\mbf{A}^T$, $\mbf{A}^H$, $\mbf{A}^{-1}$, and $a_{i,j}$ denote the transpose, conjugate transpose, inverse, and $(i,j)$-th entry of matrix $\mbf{A}$, respectively.
We use $\jmath \triangleq \sqrt{-1}$, $\otimes$, $\diag\{ \cdot \}$, $\mathbb{E} ( \cdot )$, and $\mathrm{tr} ( \cdot )$ to represent the imaginary unit, the Kronecker product, the diagonal operator, the expectation operator, and the trace of square matrix, respectively.
In addition, $\lfloor \cdot \rfloor$ returns the largest integer that is smaller than or equal to its argument, and $\bmod (a,b)$ returns the remainder of the division $a/b$.
We use $\mathbb{Z}_{+}$ to denote the set of positive integers.
The cardinality of set $\mathcal{S}$ is represented by $\left|\mathcal{S}\right|$.
Finally, the distribution of a circularly symmetric complex Gaussian (CSCG) random vector with mean $\bsm{\mu}$ and covariance matrix $\bsm{\Sigma}$ is denoted by $\mathcal{CN} (\bsm{\mu}, \bsm{\Sigma})$; and $\sim$ stands for ``distributed as''.

\section{RIS-Aided Large-Scale MIMO System} \label{section-system-model}
\subsection{System Model}
\begin{figure}
	[t]
	\centering
	\includegraphics[width=.9\columnwidth]{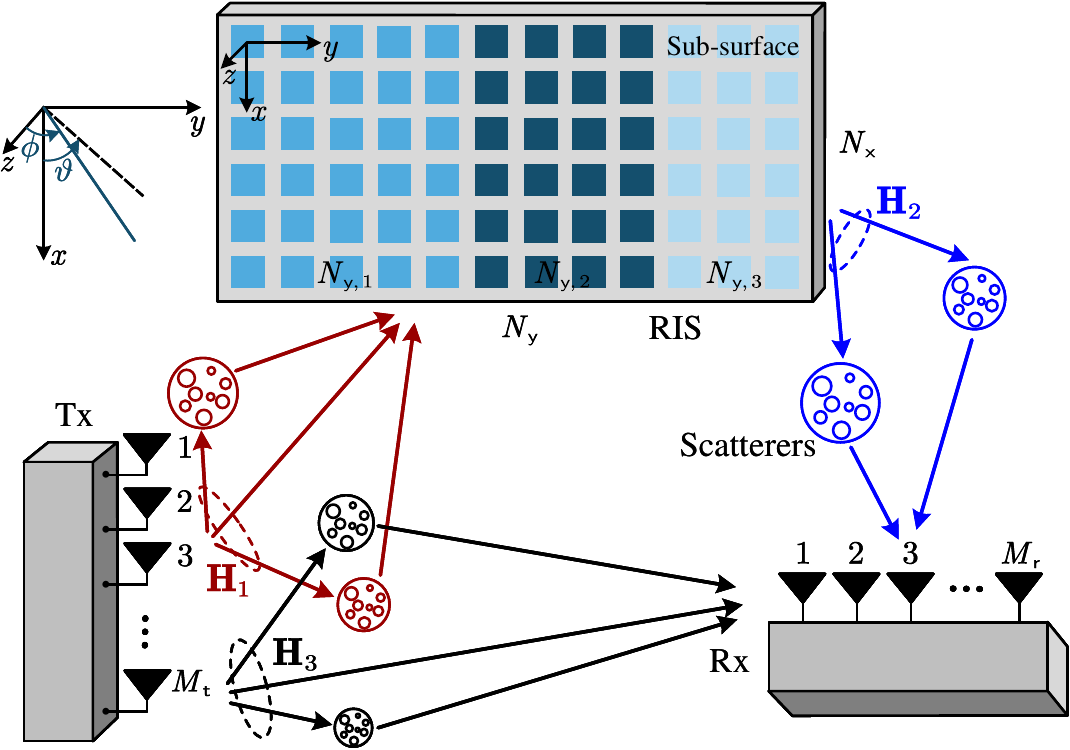}
	\caption{A RIS-aided large-scale MIMO system.}
	\label{P2P-MIMO}
\end{figure}

Consider a RIS-aided large-scale MIMO system with $M_{\sf t}$ transmit antennas and $M_{\sf r}$ receive antennas ($M_{\sf t}, M_{\sf r} \gg 1$), as illustrated in Fig. \ref{P2P-MIMO}.
We assume uniform linear arrays (ULAs) at the Tx and the Rx.
A RIS is placed in the three-dimensional (3D) Cartesian coordinate system, 
where the RIS reflecting elements are arranged in a uniform rectangular array (URA) in the $x$-$y$ plane with $N_{\sf x}$ elements in the $x$-(vertical) axis and $N_{\sf y}$ elements in the $y$-(horizontal) axis.
Following \cite{QingqingWu2019TWC,ChongwenHuang2019TWC, HuayanGuo2020WSR, XianghaoYu2019ICCC, ShuowenZhang2020JSAC,CunhuaPan2020TWC}, the magnitudes of the reflection coefficients are assumed to be a constant.\footnote{The electromagnetic (EM)-based modeling \cite{cjzhong2022elementgrouping} as well as the experimental results \cite{WankaiTang2020TWC_PathLoss} revealed that the reflection coefficients of RIS elements are related to the incident angle of the EM waves.
Extension of this work to the angle-dependent reflection model is left for our future work.}
Without loss of generality, the reflection coefficient matrix of the RIS is given by $\bsm{\Theta} = \diag \left\{  e^{\jmath \theta_{1}}, \cdots, e^{\jmath \theta_{N}} \right\} \in \mathbb{C}^{N \times N}$,
where $N = N_{\sf x} \times N_{\sf y}$ is the number of reflecting elements, and $\theta_{n}$ is the phase shift of the $n$-th reflecting element, $n \in \mathcal{N} \triangleq \left\{1,\cdots,N\right\}$.
We propose to horizontally partition the RIS into $S$ sub-surfaces,
each containing $N_s = N_{\sf x} \times N_{{\sf y}, s}$ elements, where $N_{{\sf y}, s} = t_s N_{\sf y}$ denotes the number of columns of the RIS array allocated to sub-surface $s$ with partition ratio $t_s \in [0,1]$,
$s \in \mathcal{S} \triangleq \left\{ 1,\cdots, S \right\}$.
Then, appropriate RIS partitioning can be found by optimizing $\mbf{t} = \left[t_1, \cdots, t_S\right]^T$ under the constraints $\sum_{s \in \mathcal{S}} t_s = 1$ and $N_{{\sf y}, s} \in \mathbb{Z}_{+}$, $\forall s \in \mathcal{S}$.

We assume that the Tx, the Rx, and the RIS are all deployed in the far-field region of each other.
The wireless channels are characterized by the ray-tracing based geometric channel model \cite{3GPP}.
For ease of notation, we define the following normalized steering vector as a function of angle $\phi$ and integer $M$ as
\begin{align}
	\mbf{e} \left( \phi, M \right) = \frac{1}{\sqrt{M}} \left[ 1, e^{-\jmath  \pi \phi }, \cdots, e^{-\jmath  \pi (M-1) \phi } \right]^{H} \in \mathbb{C}^{M}.
\end{align}
The array response of a ULA with $M$ elements is expressed as
\begin{align}
	\mbf{a}_M \left( \theta \right) = \mbf{e} \left( \frac{2d}{\lambda} \sin \theta, M \right),
\end{align}
where $\theta$ denotes the angle relative to the antenna boresight, $\lambda$ denotes the carrier wavelength, and $d$ stands for the element spacing between two adjacent antennas/elements.
The RIS array response is expressed as
\begin{align}
	\mbf{b}_N \left( \phi, \vartheta \right) &=  \mbf{e} \left( \frac{2d}{\lambda} \sin \phi \cos \vartheta, N_{\sf x} \right) \nonumber \\
	&~~~\otimes
	\mbf{e} \left( \frac{2d}{\lambda} \sin \phi \sin \vartheta, N_{\sf y} \right),
\end{align}
where $\left( \phi, \vartheta \right)$ is the angle of the transmitting/receiving signals defined by the spherical coordinates (as shown in Fig. \ref{P2P-MIMO}).
Based on the above notations, the baseband equivalent channel from the Tx to the RIS and from the RIS to the Rx, denoted by $\mbf{H}_1 \in \mathbb{C}^{N \times M_{\sf t}}$ and $\mbf{H}_2 \in \mathbb{C}^{M_{\sf r} \times N}$ respectively, are expressed as
\begin{align}
	\mbf{H}_1 &= \sqrt{\frac{N M_{\sf t}}{L_1}} \sum_{\ell =1}^{L_1} \alpha_\ell \mbf{b}_N \big(\phi_{\ell}^{\sf AoA}, \vartheta_{\ell}^{\sf AoA} \big) 
	\mbf{a}_{M_{\sf t}}^{H} \big(\varphi_{\ell}^{\sf AoD} \big), \label{Tx-RIS-1}  \\
	\mbf{H}_2 &= \sqrt{\frac{M_{\sf r} N}{L_2}} \sum_{\ell=1}^{L_2} \beta_{ \ell}\mbf{a}_{M_{\sf r}} \big(\varphi_{\ell}^{\sf AoA} \big) 
	\mbf{b}_{N}^{H} \big(\phi_{\ell}^{\sf AoD}, \vartheta_{\ell}^{\sf AoD} \big), \label{RIS-Rx} 
\end{align}
where $L_1$ (or $L_2$) denotes the number of resolvable paths between the Tx and the RIS (or between the RIS and the Rx), $\alpha_{\ell}$ (or $\beta_{\ell}$) denotes the complex gain of the corresponding $\ell$-th path, 
$ \left(\phi_{\ell}^{\sf AoA}, \vartheta_{\ell}^{\sf AoA} \right)$ (or $\left(\phi_{\ell}^{\sf AoD}, \vartheta_{\ell}^{\sf AoD}\right)$) denotes the $\ell$-th angle of arrival (AoA) (or angle of departure (AoD)) associated with the RIS,
and $\varphi_{\ell}^{\sf AoD}$ (or $\varphi_{\ell}^{\sf AoA}$) represents the $\ell$-th AoD (or AoA) associated with the Tx (or Rx) in the Tx-RIS (or RIS-Rx) channel.
Similarly, the channel $\mbf{H}_3 \in \mathbb{C}^{M_{\sf r} \times M_{\sf t}}$ from the Tx to the Rx is expressed as
\begin{align}
	\mbf{H}_3 = \sqrt{\frac{M_{\sf r} M_{\sf t}}{L_3}} \sum_{\ell=1}^{L_3} \gamma_{\ell} \mbf{a}_{M_{\sf r}} \big(\omega_{\ell}^{\sf AoA} \big) 
	\mbf{a}_{M_{\sf t}}^{H} \big(\omega_{\ell}^{\sf AoD} \big)
	, \label{Tx-Rx}
\end{align}
where $L_3$ denotes the number of resolvable paths between the Tx and the Rx, $\gamma_\ell$ denotes the complex gain of the corresponding $\ell$-th path, 
and $\omega_{\ell}^{\sf AoA}$ (or $\omega_{\ell}^{\sf AoD}$) represents the $\ell$-th AoA (or AoD) associated with the Rx (or Tx) in the Tx-Rx channel.
For convenience, we assume that the complex gains, $\left\{\alpha_\ell\right\}_{\ell=1}^{L_1}$, $\left\{\beta_\ell\right\}_{\ell=1}^{L_2}$, and $\left\{\gamma_\ell\right\}_{\ell=1}^{L_3}$, are all arranged in the descending order of their magnitudes.
Moreover, we assume that the angular domain information, including the AoAs, AoDs, and corresponding complex path gains are perfectly known at the Tx.
To accurately obtain the angular domain information,
the Tx (or Rx) is required to have enough spatial resolution to distinguish different Tx-RIS and Tx-Rx channel paths (or different RIS-Rx and Tx-Rx channel paths) simultaneously.
The RIS is required to have enough spatial resolution to distinguish different Tx-RIS and RIS-Rx channel paths.
Thus, we assume $L_1+L_3 \ll M_{\sf t}$, $L_2+L_3 \ll M_{\sf r}$, and $L_1, L_2 \ll N$.
This assumption is valid since only a limited number of scatterers exist in the wireless propagation environment.
In the literature, various line spectrum estimation techniques \cite{music, anm} are exploited in RIS-aided communications to estimate the angular domain information.
Please see the multiple signal classification (MUSIC) method and the estimation of signal parameters via rotational invariance technique (ESPRIT) in \cite{music}, as well as the atomic norm minimization method in \cite{anm} for more details.


Based on the system model described above, the received signal vector $\mbf{y} \in \mathbb{C}^{M_{\sf r}}$ is given by
\begin{align}
	\mbf{y} = \mbf{H}_{\rm eff} \mbf{x} + \mbf{n},
\end{align}
where 
\begin{align}
	\mbf{H}_{\rm eff} \triangleq \sqrt{\mathrm{PL}^{\sf r}} \mbf{H}_2 \bsm{\Theta} \mbf{H}_1 + \sqrt{\mathrm{PL}^{\sf d}} \mbf{H}_3
\end{align}
is the effective MIMO channel matrix from the Tx to the Rx, with $\mathrm{PL}^{\sf r}$ and $\mathrm{PL}^{\sf d}$ being the path losses of the cascaded and Tx-Rx channels, respectively,
$\mbf{x} \in \mathbb{C}^{M_{\sf t}}$ is the transmitted signal vector with zero mean, i.e., $\mathbb{E} \left[ \mbf{x} \right] = 0$,
and $\mbf{n} \sim \mathcal{CN} \left(\bsm{0}, \sigma^2 \mbf{I}_{M_{\sf r}}\right)$ denotes the independent CSCG noise vector at the Rx, with $\sigma^2$ being the average noise power.
Moreover, the transmit signal covariance matrix is defined as $\mbf{Q} \triangleq \mathbb{E} \left[ \mbf{x} \mbf{x}^{H} \right] \in \mathbb{C}^{M_{\sf t} \times M_{\sf t}}$, where $\mbf{Q} \succeq \mbf{0}$.
The power budget is denoted by $P$, i.e., $\mathbb{E} \left[ \left\| \mbf{x} \right\|^2 \right] \leq P$, or equivalently, $\mathrm{tr} \left(\mbf{Q}\right) \leq P$.

\subsection{Phase-Shift Structure Specification}
\begin{figure}
	[t]
	\centering
	\includegraphics[width=.83\columnwidth]{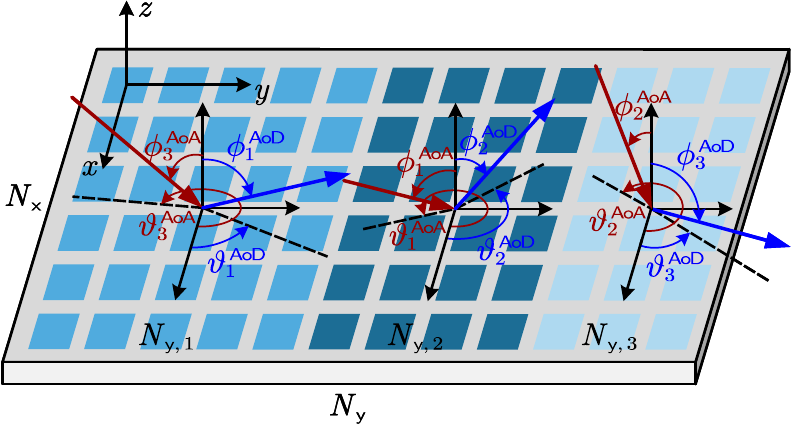}
	\caption{An illustration of the RIS partitioning design, where the RIS is partitioned into three sub-surfaces with sub-surface 1 serving the AoA $\left(\phi_{3}^{\sf AoA}, \vartheta_{3}^{\sf AoA}\right)$ and the AoD $\left(\phi_{1}^{\sf AoD}, \vartheta_{1}^{\sf AoD}\right)$, 
	sub-surface 2 serving the AoA $\left(\phi_{1}^{\sf AoA}, \vartheta_{1}^{\sf AoA}\right)$ and the AoD $\left(\phi_{2}^{\sf AoD}, \vartheta_{2}^{\sf AoD}\right)$,
	and sub-surface 3 serving the AoA $\left(\phi_{2}^{\sf AoA}, \vartheta_{2}^{\sf AoA}\right)$ and the AoD $\left(\phi_{3}^{\sf AoD}, \vartheta_{3}^{\sf AoD}\right)$.}
	\label{RIS_Partitioning}
\end{figure}
The RIS partitioning design provides a new paradigm that treats the sub-surfaces instead of individual elements as design entities, 
thereby reducing the dimension of the optimization space involved in the passive beamforming design.
In this paper, we propose to design each sub-surface to reflect the incident signals from an AoA of the Tx-RIS channel to an AoD of the RIS-Rx channel, referred to as \textit{anomalous reflection}.
Fig. \ref{RIS_Partitioning} illustrates the basic idea of the proposed design where the RIS is partitioned into three sub-surfaces, 
with sub-surface 1 reflecting the signals from the AoA $\left(\phi_{3}^{\sf AoA}, \vartheta_{3}^{\sf AoA}\right)$ to the AoD $\left(\phi_{1}^{\sf AoD}, \vartheta_{1}^{\sf AoD}\right)$, 
sub-surface 2 reflecting the signals from the AoA $\left(\phi_{1}^{\sf AoA}, \vartheta_{1}^{\sf AoA}\right)$ to the AoD $\left(\phi_{2}^{\sf AoD}, \vartheta_{2}^{\sf AoD}\right)$,
and sub-surface 3 reflecting the signals from the AoA $\left(\phi_{2}^{\sf AoA}, \vartheta_{2}^{\sf AoA}\right)$ to the AoD $\left(\phi_{3}^{\sf AoD}, \vartheta_{3}^{\sf AoD}\right)$.
According to the generalized Snell’s law \cite{Nanfang2011Snell}, anomalous reflection can be achieved by setting a linear variation of the phase shifts across the $x$-$y$ plane.
Specifically, for sub-surface $s$ to achieve the anomalous reflection from the incident angle $\left(\phi_{u}^{\sf AoA}, \vartheta_{u}^{\sf AoA}\right)$ to the reflection angle $\left(\phi_{v}^{\sf AoD}, \vartheta_{v}^{\sf AoD}\right)$, 
the phase difference between any two adjacent elements is set to $\frac{2\pi d}{\lambda}g_{{\sf x},s}$ along the $x$-axis, and to $\frac{2\pi d}{\lambda}g_{{\sf y},s}$ along the $y$-axis, respectively, where
\begin{align} \label{phase-gradient}
	\mbf{g}_s \triangleq
	\begin{bmatrix} g_{{\sf x},s}  \\  g_{{\sf y},s} \end{bmatrix}
	=
	\begin{bmatrix} \sin\phi_{v}^{\sf AoD} \cos\vartheta_{v}^{\sf AoD} - \sin\phi_{u}^{\sf AoA} \cos\vartheta_{u}^{\sf AoA} \\  \sin\phi_{v}^{\sf AoD} \sin\vartheta_{v}^{\sf AoD} - \sin\phi_{u}^{\sf AoA} \sin\vartheta_{u}^{\sf AoA} \end{bmatrix}
\end{align}
is referred to as the \textit{phase gradient} of sub-surface $s$, with $g_{{\sf x},s}$ and $g_{{\sf y},s}$ being the phase gradient components along the $x$- and $y$-axes, respectively, $u \in \mathcal{L}_1 \triangleq \left\{1,\cdots, L_1\right\}$, and $v \in \mathcal{L}_2 \triangleq \left\{1,\cdots, L_2\right\}$.
The feasible set of the phase gradient $\mbf{g}_s$ is expressed as
\begin{align} \label{discrete}
	\mathcal{F} \triangleq \left\{ \left.  [\zeta_{{\sf x}, u,v}, \zeta_{{\sf y}, u,v} ]^T
	\right|
	u \in \mathcal{L}_1, v \in \mathcal{L}_2
	\right\},
\end{align}
where $\zeta_{{\sf x}, u,v} \triangleq \sin\phi_{v}^{\sf AoD} \cos\vartheta_{v}^{\sf AoD} - \sin\phi_{u}^{\sf AoA} \cos\vartheta_{u}^{\sf AoA}$, and $\zeta_{{\sf y}, u,v} \triangleq \sin\phi_{v}^{\sf AoD} \sin\vartheta_{v}^{\sf AoD} - \sin\phi_{u}^{\sf AoA} \sin\vartheta_{u}^{\sf AoA}$.\footnote{
	We assume that
	$[\zeta_{{\sf x}, u,v}, \zeta_{{\sf y}, u,v} ]^T$ has distinct values for different $u$ and/or $v$. 
	That is, each element in $\mathcal{F}$ is unique.}

\begin{remk}
	Although it is crucial for the sub-surfaces to reflect signals along the strong Tx-RIS-Rx paths in order to ensure sufficient link budget,
	it is unwise to let the sub-surfaces select the strongest Tx-RIS-Rx path all together, i.e., set $\mbf{g}_s = [\zeta_{{\sf x}, 1,1}, \zeta_{{\sf y}, 1,1} ]^T$ for all $s \in \mathcal{S}$.
	The reason is that the constructed cascaded channel is rank-deficient and thus cannot support multiplexing.
\end{remk}

Apart from the phase gradient, a \textit{common phase shift}, denoted as $\psi_s$, is imposed on every element of sub-surface $s$, $s\in \mathcal{S}$.
That is, the reflection coefficient of the $n$-th element is given by
\begin{align} \label{structured}
	e^{\jmath \theta_{n}} = e^{\jmath \psi_s}
	e^{\jmath \frac{2\pi (n_{\sf x}-1) d}{\lambda} g_{{\sf x}, s}}
	e^{\jmath \frac{2\pi (n_{\sf y}-1) d}{\lambda} g_{{\sf y}, s}},
	~~n \in \mathcal{N},
\end{align}
where $n_{\sf x} = \lfloor (n-1) / N_{\sf x}\rfloor + 1$, $n_{\sf y} = \bmod (n-1, N_{\sf x}) +1 $, and $s$ is the sub-surface index of the $n$-th reflecting element.
In \eqref{structured}, the common phase shift $\psi_s$ adjusts the reflected wavefront, and the phase gradient $\mbf{g}_s$ determines the direction of the beam reflected by sub-surface $s$.

\subsection{Problem Formulation} 
Based on the above RIS partitioning design, 
the total number of variables for passive beamforming optimization is reduced from $N$ to $4S$,
including the RIS partitioning vector $\mbf{t} = \left[t_1, \cdots, t_S\right]^T \in \mathbb{R}^{S}$, the phase gradients $\mbf{G} = \left[\mbf{g}_1, \cdots, \mbf{g}_S\right] \in \mathbb{R}^{2\times S}$, and the common phase shifts $\bsm{\psi} = \left[\psi_1, \cdots, \psi_S\right]^T\in \mathbb{R}^{S}$,
where $S$ is assumed to be given.
We aim to maximize the RIS-aided MIMO channel capacity by jointly optimizing the transmit covariance matrix $\mbf{Q}$ and the RIS reflection matrix $\bsm{\Theta}$, subject to the total power constraint at the Tx and the constraints introduced by RIS partitioning.
This problem is formulated as
\begin{align}
	\textrm{(P1):} ~\max_{\mathbf{Q},\mathbf{\Theta}\left(\mbf{t}, \mbf{G}, \bsm{\psi}\right) } \quad & \log \det \left( \mathbf{I}_{M_{\sf r}} + \frac{1}{\sigma^2} \mathbf{H}_{\rm eff} \mathbf{Q} \mathbf{H}_{\rm eff}^{H} \right) \nonumber \\ 
	\operatorname{ s.t. } \quad
	&{\textrm{C1:}} ~ \mathrm{tr} \left(\mathbf{Q}\right) \leq P, \nonumber \\
	&{\textrm{C2:}} ~ \mathbf{Q} \succeq \mathbf{0}, \nonumber \\
	& {\textrm{C3:}} ~ \sum_{s \in \mathcal{S}} t_s = 1, \nonumber \\
	&{\textrm{C4:}} ~t_s N_{\sf y} \in \mathbb{Z}_{+}, ~~\forall s \in \mathcal{S}, \nonumber \\
	&{\textrm{C5:}} ~ \mbf{g}_s \in \mathcal{F}, ~~\forall s \in \mathcal{S}, \nonumber \\
	&{\textrm{C6:}} ~ \psi_s \in \left[ 0, 2\pi \right), ~~\forall s \in \mathcal{S}.  \nonumber 
\end{align}
The difficulty for solving (P1) is twofold.
Firstly, due to C4, the optimization w.r.t. $\mbf{t}$ is an \textit{integer programming} problem that is known to be NP-complete \cite{schrijver1998theory}. 
Secondly, C5 restricts the phase gradients to discrete values, which makes the problem even harder.
As such, the RIS partitioning based system design seems to complicate the beamforming design problem.
However, we next show that (P1) has a rather simple expression in the asymptotic regime, and the asymptotic solution can serve as a guideline for the finite-size system design.

\section{Asymptotic Expression of (P1)} \label{sec-asymptotic-formulation}
RIS-assisted MIMO systems in next-generation wireless networks are foreseeable to support a large number of antennas at the Tx and Rx, as well as a very large number of reflecting elements at the RIS.
This inspires us to study the formulation of (P1) in the asymptotic regime where $M_{\sf t}$, $M_{\sf r}$, $N_{\sf x}$, and $N_{\sf y}$ go to infinity, i.e., 
\begin{align}
	M_{\sf t}, M_{\sf r}, N_{\sf x}, N_{\sf y} \rightarrow \infty. \label{asymptotic-regime}
\end{align}
To this end, we rewrite the cascaded channel as
\begin{align}
	&\sqrt{\mathrm{PL}^{\sf r}} \mbf{H}_2 \bsm{\Theta} \mbf{H}_1 \nonumber \\
	=& \sqrt{\frac{\mathrm{PL}^{\sf r} M_{\sf r} M_{\sf t}}{L_1 L_2}} N \left( \sum_{v=1}^{L_2} \beta_{v}\mbf{a}_{M_{\sf r}} \big(\varphi_{v}^{\sf AoA} \big) 
	\mbf{b}_{N}^{H} \big(\phi_{v}^{\sf AoD}, \vartheta_{v}^{\sf AoD} \big) \right)
	  \nonumber \\
	&\bsm{\Theta} \left(
	\sum_{u=1}^{L_1} \alpha_u \mbf{b}_N \big(\phi_{u}^{\sf AoA}, \vartheta_{u}^{\sf AoA} \big) 
	\mbf{a}_{M_{\sf t}}^{H} \big(\varphi_{u}^{\sf AoD} \big) \right) \nonumber \\
	=& \sqrt{\frac{\mathrm{PL}^{\sf r} M_{\sf r} M_{\sf t}}{L_1 L_2}} N \sum_{u=1}^{L_1} \sum_{v=1}^{L_2} \alpha_u \beta_v \mbf{a}_{M_{\sf r}} \big(\varphi_{v}^{\sf AoA} \big) \nonumber \\
	&\underbrace{ \mbf{b}_{N}^{H} \big(\phi_{v}^{\sf AoD}, \vartheta_{v}^{\sf AoD} \big) \bsm{\Theta} \mbf{b}_N \big(\phi_{u}^{\sf AoA}, \vartheta_{u}^{\sf AoA} \big)  }_{d_{u,v}}
	\mbf{a}_{M_{\sf t}}^{H} \big(\varphi_{u}^{\sf AoD} \big)  \nonumber \\
	=& \sqrt{\frac{\mathrm{PL}^{\sf r} M_{\sf r} M_{\sf t}}{L_1 L_2}} N \sum_{u=1}^{L_1} \sum_{v=1}^{L_2} \alpha_u \beta_v d_{u,v} \mbf{a}_{M_{\sf r}} \big(\varphi_{v}^{\sf AoA} \big) \mbf{a}_{M_{\sf t}}^{H} \big(\varphi_{u}^{\sf AoD} \big) \label{sum-path},
\end{align}
where
\begin{align} \label{passive-beamforming-gain}
	d_{u,v} = \frac{1}{N} \sum_{n_{\sf x}=1}^{N_{\sf x}} \sum_{n_{\sf y}=1}^{N_{\sf y}} 
	e^{\jmath \theta_{n} }
	e^{-\jmath k
		\left( \left( n_{\sf x}-1 \right) \zeta_{{\sf x},u,v} +
		\left( n_{\sf y}-1 \right) \zeta_{{\sf y},u,v} \right)}
\end{align}
is referred to as the \textit{normalized passive beamforming gain} associated with the $u$-th path from the Tx to the RIS and the $v$-th path from the RIS to the Rx,
with $k = \frac{2 \pi d}{\lambda}$ and $n=  n_{\sf x}+(n_{\sf y}-1)N_{\sf x}$.
From \eqref{sum-path}, we observe that the power gain of the $(u, v)$-th Tx-RIS-Rx path is determined not only by the path gains of the individual Tx-RIS and RIS-Rx channels, i.e., $\alpha_u$ and $\beta_v$, but also by the normalized passive beamforming gain $d_{u,v}$ that can be adjusted by controlling the phase shifts of the reflecting elements. 
For notation simplicity, we define $\eta_{{\sf x},s,u,v} \triangleq g_{{\sf x},s} -\zeta_{{\sf x},u,v}$ and $\eta_{{\sf y}, s,u,v} \triangleq g_{{\sf y},s} - \zeta_{{\sf y},u,v}$.
Substituting \eqref{structured} into \eqref{passive-beamforming-gain} yields
\begin{align}
	d_{u,v} &= \frac{1}{N}\sum_{s \in \mathcal{S}} e^{\jmath \psi_s}
	\left(
	\sum_{n_{\sf x}=1}^{N_{\sf x}} e^{\jmath k (n_{\sf x}-1) \eta_{{\sf x},s,u,v}} \right) \nonumber \\
	&~~~\left(
	\sum_{n_{\sf y}=N_{{\sf y}, s-1}^{\sf tot} +1}^{N_{{\sf y}, s}^{\sf tot}} e^{\jmath k (n_{\sf y}-1) \eta_{{\sf y}, s,u,v}}
	\right) \nonumber \\
	&= \sum_{s \in \mathcal{S}} e^{\jmath \psi_s}
	\left( e^{\jmath \frac{k}{2} (N_{\sf x}-1) \eta_{{\sf x},s,u,v} }  \frac{\mathrm{sinc} \left(\frac{k}{2} N_{\sf x}\eta_{{\sf x},s,u,v} \right)}{\mathrm{sinc} \left( \frac{k}{2} \eta_{{\sf x},s,u,v} \right)} \right)  \nonumber \\
	&~~~\left( e^{\jmath \frac{k}{2} (N_{{\sf y}, s}^{\sf tot}+N_{{\sf y}, s-1}^{\sf tot}-1) \eta_{{\sf y}, s,u,v} } t_s \frac{\mathrm{sinc} \left(\frac{k}{2} t_s N_{\sf y}\eta_{{\sf y}, s,u,v}\right)}{\mathrm{sinc} \left( \frac{k}{2} \eta_{{\sf y}, s,u,v} \right)} \right) \nonumber \\
	&= \sum_{s \in \mathcal{S}} e^{\jmath \widetilde{\psi}_s} t_s  \frac{\mathrm{sinc} \left(\frac{k}{2} N_{\sf x} \eta_{{\sf x},s,u,v}\right)}{\mathrm{sinc} \left( \frac{k}{2} \eta_{{\sf x},s,u,v} \right)} \frac{\mathrm{sinc} \left(\frac{k}{2} t_s N_{\sf y} \eta_{{\sf y}, s,u,v}\right)}{\mathrm{sinc} \left( \frac{k}{2} \eta_{{\sf y}, s,u,v} \right)} , \label{sum-tile-gain}
\end{align}
where
\begin{align} \label{widetile-crf}
	\widetilde{\psi}_s &= \psi_s + \frac{k}{2} (N_{\sf x}-1) \eta_{{\sf x},s,u,v} \nonumber \\
	&~~~+ \frac{k}{2} (N_{{\sf y}, s}^{\sf tot}+N_{{\sf y}, s-1}^{\sf tot}-1) \eta_{{\sf y}, s,u,v},
\end{align} 
with $N_{{\sf y}, s}^{\sf tot} \triangleq \sum_{i=1}^{s} N_{{\sf y}, i}$ denoting the total number of columns of the first $s$ sub-surface(s), $s \in \mathcal{S}$, and $N_{{\sf y}, 0}^{\sf tot} \triangleq 0$.
Eq. \eqref{sum-tile-gain} represents $d_{u,v}$ as the summation of the passive beamforming gain brought by each sub-surface.
The expression in \eqref{sum-tile-gain} is still complicated due to the involvement of sinc functions.

We now introduce the asymptotic condition $N_{\sf x}, N_{\sf y} \rightarrow \infty$ to simplify $d_{u,v}$ as
\begin{align} \label{infty-duv}
	\lim_{N_{\sf x} ,N_{\sf y} \rightarrow \infty} d_{u,v} = 
	\sum_{s \in \mathcal{S}}  \mathbbm{1}\{\eta_{{\sf x},s,u,v}, \eta_{{\sf y},s,u,v}\} e^{\jmath \widetilde{\psi}_s} t_s,
\end{align}
where the indicator function $\mathbbm{1}\{a, b\}$ is defined as
\begin{align} \label{indicator-func}
	\mathbbm{1}\{a, b\} = 
	\begin{cases}
		1, & a = b = 0, \\
		0, & \text{otherwise}.
	\end{cases}
\end{align}
The condition $\eta_{{\sf x},s,u,v} = \eta_{{\sf y}, s,u,v} = 0$ means that the phase gradient $\mbf{g}_s$ of sub-surface $s$ is designed to reflect the signals from the AoA $\left(\phi_{u}^{\sf AoA}, \vartheta_{u}^{\sf AoA}\right)$ to the AoD $\left(\phi_{v}^{\sf AoD}, \vartheta_{v}^{\sf AoD}\right)$, or equivalently, to align the $u$-th path of the Tx-RIS channel with the $v$-th path of the RIS-Rx channel.
Eq. \eqref{infty-duv} shows that when the RIS is infinitely large, only the sub-surfaces that exactly align $\left(\phi_{u}^{\sf AoA}, \vartheta_{u}^{\sf AoA}\right)$ with $\left(\phi_{v}^{\sf AoD}, \vartheta_{v}^{\sf AoD}\right)$ contribute to the passive beamforming gain $d_{u,v}$ of the $(u,v)$-th Tx-RIS-Rx path.

In the following, we show that in the considered asymptotic regime, restricting each Tx-RIS path and each RIS-Rx path to be served by at most one sub-surface does not lose the optimality of (P1).
\begin{itemize}	
	\item With $M_{\sf t}, M_{\sf r} \rightarrow \infty$, both the Tx and the Rx have infinite spatial resolution to create orthogonal sub-channels for different paths.
	If there exists more than one sub-surface serving the same Tx-RIS-Rx path, 
	we can always adjust the corresponding common phase shifts to maximize the received power through this path and thus improve the achievable rate. 
	This adjustment is equivalent to merging the involved sub-surfaces into a single sub-surface.
	
	\item If there exists two sub-surfaces serving a common Tx-RIS (or RIS-Rx) path but different RIS-Rx (or Tx-RIS) paths, the cascaded channel constructed by the two sub-surfaces is rank-one and thus cannot support multiplexing.
	Letting both the two sub-surfaces align with the stronger RIS-Rx (or Tx-RIS) path results in a higher received power, and hence achieves a higher data rate.
	Thus, the two sub-surfaces can be merged into one sub-surface.
\end{itemize}
Therefore,
the asymptotically optimal solution can be achieved with at most $\min \{L_1, L_2\}$ sub-surfaces.
\begin{remk} \label{remk2}
	Denote the minimum $S$ required to achieve the asymptotically optimal solution as $S^\star_{\min}$, and we have $S^\star_{\min} \leq \min \{L_1, L_2\}$.
	For the case of $S = S^\star_{\min}$, the asymptotically optimal solution is achieved with arbitrary common phase shifts of the sub-surfaces, since the served Tx-RIS-Rx path pairs are distinct and asymptotically orthogonal to each other.
	For the case of $S > S^\star_{\min}$, the asymptotically optimal solution is achieved when some sub-surfaces serving the same Tx-RIS-Rx path pair, and the common phase shifts are properly adjusted to virtually merge these sub-surfaces into a single sub-surface.
	This, however, results in a larger optimization space and the adjustment of common phase shifts incurs additional computational cost.
	Therefore, we are interested in $S^\star_{\min}$ which achieves the asymptotic optimum with lowest computational complexity.
\end{remk}

Based on the above,
we restrict each Tx-RIS path and each RIS-Rx path to be served by at most one sub-surface, and therefore, $S \leq \min \{L_1, L_2\}$.
For any $(u,v)$, there is at most one term in the summation of \eqref{infty-duv} being non-zero.
Consequently,
the optimization of the phase gradients $\mbf{G}$ reduces to the \textit{tripartite matching} of the Tx-RIS paths, the sub-surfaces, and the RIS-Rx paths.
From \eqref{infty-duv}, the passive beamforming gain of a Tx-RIS-Rx path is determined by the partition ratio $t_s$ and the common phase shift $\psi_s$ of the serving sub-surface $s$, and is irrelevant to the specific position of sub-surface $s$ on the RIS.
This implies that the specific position of a sub-surface on the RIS does not affect the performance in the asymptotic regime.
Thus, the tripartite matching problem reduces to the \textit{bipartite matching} between the Tx-RIS paths and the RIS-Rx paths.
As such, we define a path pairing matrix $\mbf{B} \in \mathbb{C}^{L_1 \times L_2}$ with $b_{u, v}$ being the $(u,v)$-th entry of $\mbf{B}$.
When there exists a sub-surface to align the $u$-th Tx-RIS path to the $v$-th RIS-Rx path, the corresponding entry of $\mbf{B}$ is given by $b_{u,v} = 1$; otherwise $b_{u,v} = 0$.
The path pairing constraint is expressed as
\begin{align}
	{ {\textrm{C7:}} ~ } 
	\begin{cases}
		\sum_{u=1}^{L_1} b_{u,v} \leq 1,~~ \forall v, \nonumber \\
		\sum_{v=1}^{L_2} b_{u,v} \leq 1,~~ \forall u, \nonumber \\
		b_{u,v} \in \left\{0,1\right\}, ~~ \forall u,v, \nonumber \\
		\sum_{u=1}^{L_1} \sum_{v=1}^{L_2} b_{u,v} = S. \nonumber
	\end{cases}
\end{align}

The effective channel $\mbf{H}_{\rm eff}$ can be rewritten as
\begin{align} \label{matrix-plus} 
	\mbf{H}_{\rm eff}  &= \sqrt{\frac{\mathrm{PL}^{\sf r} M_{\sf r} M_{\sf t}}{L_1 L_2}} N \mbf{A}_{M_{\sf r}} \left( \bsm{\varphi}^{\sf AoA} \right) \bsm{\Sigma} \mbf{A}_{M_{\sf t}}^{H} \left( \bsm{\varphi}^{\sf AoD} \right) \nonumber \\
	&~~~+ \sqrt{\frac{\mathrm{PL}^{\sf d} M_{\sf r} M_{\sf t}}{L_3}}  \mbf{A}_{M_{\sf r}} \left( \bsm{\omega}^{\sf AoA} \right) \diag \left(\bsm{\gamma}\right) \mbf{A}_{M_{\sf t}}^{H} \left( \bsm{\omega}^{\sf AoD} \right) ,
\end{align}
where 
$\mbf{A}_{M_{\sf r}} \left( \bsm{\varphi}^{\sf AoA} \right) \triangleq  \left[ \mbf{a}_{M_{\sf r}} \left(\varphi_{1}^{\sf AoA} \right), \cdots, \mbf{a}_{M_{\sf r}} \left(\varphi_{L_2}^{\sf AoA} \right) \right] \in \mathbb{C}^{M_{\sf r} \times L_2}$ collects the $L_2$ arrival steering vectors of the RIS-Rx channel. 
Similar definitions are applied to $\mbf{A}_{M_{\sf t}} \left( \bsm{\varphi}^{\sf AoD} \right)$, $\mbf{A}_{M_{\sf r}} \left( \bsm{\omega}^{\sf AoA} \right)$, and $\mbf{A}_{M_{\sf t}} \left( \bsm{\omega}^{\sf AoD} \right)$, i.e., 
$\mbf{A}_{M_{\sf t}} \left( \bsm{\varphi}^{\sf AoD} \right) \triangleq  \left[ \mbf{a}_{M_{\sf t}} \left(\varphi_{1}^{\sf AoD} \right), \cdots, \mbf{a}_{M_{\sf t}} \left(\varphi_{L_1}^{\sf AoD} \right) \right] \in \mathbb{C}^{M_{\sf t} \times L_1}$,
$\mbf{A}_{M_{\sf r}} \left( \bsm{\omega}^{\sf AoA} \right) \triangleq  \left[ \mbf{a}_{M_{\sf r}} \left(\omega_{1}^{\sf AoA} \right), \cdots, \mbf{a}_{M_{\sf r}} \left(\omega_{L_3}^{\sf AoA} \right) \right] \in \mathbb{C}^{M_{\sf r} \times L_3}$, and
$\mbf{A}_{M_{\sf t}} \left( \bsm{\omega}^{\sf AoD} \right) \triangleq \left[ \mbf{a}_{M_{\sf t}} \left(\omega_{1}^{\sf AoD} \right), \cdots,\mbf{a}_{M_{\sf t}} \left(\omega_{L_3}^{\sf AoD} \right) \right] \in \mathbb{C}^{M_{\sf t} \times L_3}$.
The $(v,u)$-th entry of $\bsm{\Sigma} \in \mathbb{C}^{L_2 \times L_1}$ is given by $\sigma_{v,u} = \alpha_u \beta_v d_{u,v}$.
We notice that $\bsm{\Sigma}$ has $S$ non-zero entries in total, with at most one non-zero entry in each row/column.
Hence, we delete the steering vectors in $\mbf{A}_{M_{\sf r}} \left( \bsm{\varphi}^{\sf AoA} \right)$ and $\mbf{A}_{M_{\sf t}} \left( \bsm{\varphi}^{\sf AoD} \right)$ corresponding to the paths that are not served by any sub-surfaces, and rearrange the order of the remaining ones such that $\bsm{\Sigma}$ becomes a diagonal matrix $\widetilde{\bsm{\Sigma}}$.
Then, the first term of \eqref{matrix-plus} can be equivalently expressed as $\sqrt{\frac{\mathrm{PL}^{\sf r} M_{\sf r} M_{\sf t}}{L_1 L_2}} N \widetilde{\mbf{A}}_{M_{\sf r}} \left( \bsm{\varphi}^{\sf AoA} \right) \widetilde{\bsm{\Sigma}} \widetilde{\mbf{A}}_{M_{\sf t}}^{H} \left( \bsm{\varphi}^{\sf AoD} \right)$,
where $\widetilde{\mbf{A}}_{M_{\sf r}} \left( \bsm{\varphi}^{\sf AoA} \right)$ and $\widetilde{\mbf{A}}_{M_{\sf t}} \left( \bsm{\varphi}^{\sf AoD} \right)$ are respectively the rearranged forms of $\mbf{A}_{M_{\sf r}} \left( \bsm{\varphi}^{\sf AoA} \right)$ and $\mbf{A}_{M_{\sf t}} \left( \bsm{\varphi}^{\sf AoD} \right)$.
Define $\mbf{A}_{M_{\sf r}} \triangleq \left[ \widetilde{\mbf{A}}_{M_{\sf r}} \left( \bsm{\varphi}^{\sf AoA} \right), \mbf{A}_{M_{\sf r}} \left( \bsm{\omega}^{\sf AoA} \right) \right]$ and $\mbf{A}_{M_{\sf t}} \triangleq$ $\left[ \widetilde{\mbf{A}}_{M_{\sf t}} \left( \bsm{\varphi}^{\sf AoD} \right), \mbf{A}_{M_{\sf t}} \left( \bsm{\omega}^{\sf AoD} \right) \right]$,
we recast \eqref{matrix-plus} more compactly as 
\begin{align} \label{truncated-svd}
	\mbf{H}_{\rm eff} = \mbf{A}_{M_{\sf r}}  \bsm{\Sigma}_{\rm eff} \mbf{A}_{M_{\sf t}}^{H},
\end{align}
where
\begin{align}
	\bsm{\Sigma}_{\rm eff} = 
	\begin{bmatrix} \sqrt{\frac{\mathrm{PL}^{\sf r} M_{\sf r} M_{\sf t}}{L_1 L_2}} N  \widetilde{\bsm{\Sigma}}  & \bsm{0} \\
		\bsm{0}  & \sqrt{\frac{\mathrm{PL}^{\sf d} M_{\sf r} M_{\sf t}}{L_3}} \diag \left(\bsm{\gamma}\right)  \\ \end{bmatrix}.
\end{align}
Eq. \eqref{truncated-svd} can be regarded as the virtual channel representation \cite{virtual_channel_representation} of the effective channel $\mbf{H}_{\mathrm{eff}}$ by viewing $\mbf{H}_{\mathrm{eff}}$ in beamspace.
When $M_{\sf t}, M_{\sf r} \rightarrow \infty$,
we can readily show that the array response vectors at the Tx and the Rx are asymptotically orthogonal, i.e., $\mbf{A}_{M_{\sf t}}^{H} \mbf{A}_{M_{\sf t}} \rightarrow \mbf{I}$ and $\mbf{A}_{M_{\sf r}}^{H} \mbf{A}_{M_{\sf r}} \rightarrow \mbf{I}$.
Therefore, \eqref{truncated-svd} can be approximated as the truncated singular value decomposition (SVD) of $\mbf{H}_{\rm eff}$.

With the above asymptotic SVD representation of $\mbf{H}_{\rm eff}$,
the optimal transmit covariance matrix $\mbf{Q}$ of (P1) is given by the \textit{eigenmode transmission} \cite{Goldsmith2005Wireless}:
\begin{align}
	\mbf{Q}^{\star} = \mbf{A}_{M_{\sf t}} \diag \left\{ \mbf{p}\right\} \mbf{A}_{M_{\sf t}}^{H},
\end{align}
where $\mbf{p} = [(\mbf{p}^{\sf r})^T, (\mbf{p}^{\sf d})^T ]^T = [p_1^{\sf r}, \cdots, p_S^{\sf r}, p_1^{\sf d} ,\cdots,  p_{L_3}^{\sf d}]^T$ with $p_s^{\sf r}$ being the transmit power allocated to the $s$-th Tx-RIS-Rx path, and $p_i^{\sf d}$ being the transmit power allocated to the $i$-th Tx-Rx path.
Substituting $\mbf{Q}^{\star}$ into the objective function of (P1), we obtain 
\begin{align} \label{capacity}
	C \left(\mbf{p}, \mbf{t}, \mbf{B}\right) &\triangleq \sum_{s\in \mathcal{S}} \log\left(1+ m_s^{\sf r} p_s^{\sf r} t_s^2  \right) \nonumber \\
	&~~~+  \sum_{i\in \mathcal{L}_3} \log\left(1+ m_{i}^{\sf d} p_{i}^{\sf d}  \right),
\end{align}
where $m_s^{\sf r} \triangleq \mathrm{PL}^{\sf r}\frac{ M_{\sf t} M_{\sf r} N^2 \left| \alpha_{u_s} \beta_{v_s}\right| ^2 }{L_1 L_2\sigma^2}$ with $u_s$ and $v_s$ being the indices of the Tx-RIS and RIS-Rx paths served by sub-surface $s$, respectively, $m_i^{\sf d} \triangleq\mathrm{PL}^{\sf d} \frac{M_{\sf t} M_{\sf r} \left|\gamma_i\right|^2}{L_3 \sigma^2}$, and $\mathcal{L}_3 \triangleq \left\{1,\cdots, L_3\right\}$.
The dependence of the objective on the path pairing matrix $\mbf{B}$ is shown implicitly in the first term of \eqref{capacity}, i.e., how the indices $u_s$ and $v_s$ are associated in $m_s^{\sf r}$.
Note that the coefficients $m_s^{\sf r}$ and $m_i^{\sf d}$ are unbounded if $M_{\sf t}, M_{\sf r}, N_{\sf x}, N_{\sf y} \rightarrow \infty$.
In practice, the values of $m_s^{\sf r}$ and $m_i^{\sf d}$ are relatively small since $\mathrm{PL}^{\sf r}$ and $\mathrm{PL}^{\sf d}$ are typically in the order of $10^{-12} \sim 10^{-16}$ and $10^{-6} \sim 10^{-8}$, respectively \cite{WankaiTang2020TWC_PathLoss, emil2020physics_wcl}, while $M_{\sf t}$ and $M_{\sf r}$ are in the order of hundreds and $N$ is in the order of thousands.
Thus, we keep $m_s^{\sf r}$ and $m_i^{\sf d}$ as finite-valued parameters in \eqref{capacity}.

Based on the above, (P1) has the following asymptotic form:
\begin{align}
	\textrm{(P2):} ~~\max_{\mathbf{p},\mbf{t}, \mbf{B}} \quad & C \left(\mbf{p}, \mbf{t}, \mbf{B}\right) \nonumber \\ 
	\operatorname{ s.t. } \quad
	& {\textrm{C8:}}~  p_s^{\sf r} \geq 0, ~~\forall s \in \mathcal{S}, \nonumber \\
	& {\textrm{C9:}}~  p_i^{\sf d} \geq 0, ~~\forall i \in \mathcal{L}_3, \nonumber \\
	& {\textrm{C10:}}~ \sum_{s\in \mathcal{S}} p_s^{\sf r} + \sum_{i\in \mathcal{L}_3} p_i^{\sf d} = P, \nonumber \\
	& {\textrm{C11:}}~ t_s \geq 0, ~~\forall s \in \mathcal{S}, \nonumber \\
	&{\textrm{C3}}, {\textrm{C7}}. \nonumber
\end{align}
(P2) is a mixed-integer program, and the objective function is non-convex over the RIS partitioning vector $\mbf{t}$.
Notably, in (P2), the number $S$ of sub-surfaces is no longer predetermined.
Instead, since we restrict each sub-surface to serve a distinct Tx-RIS-Rx path pair,
the optimal $S$ of (P2) is exactly $S^\star_{\min}$.

The asymptotic analysis in this section can be generalized to more arbitrary 2D shapes of the sub-surfaces.
We leave the discussion to Appendix A.

\section{Optimal Solution to (P2)} \label{sec-algorithm}
In this section, we first study the problem of optimizing the power allocation $\mbf{p}$ and the RIS partitioning $\mbf{t}$ for any fixed path pairing matrix $\mbf{B}$:
\begin{align}
	\textrm{(P3):}~~\max_{ \mbf{p}, \mbf{t} } \quad &  C \left(\mbf{p},\mbf{t}\right) \nonumber \\ 
	\operatorname{ s.t. } \quad
	& \textrm{C8}, \textrm{C9}, \textrm{C10}, \textrm{C11}, \textrm{C3} , \nonumber
\end{align}
where $C(\mbf{p}, \mbf{t})$ is an abbreviation of $C(\mbf{p}, \mbf{t}, \mbf{B})$ for fixed $\mbf{B}$.
Then, we derive the optimal $\mbf{B}$ in Section \ref{opt-pairing}, which is irrelevant to the choices of $\mbf{p}$ and $\mbf{t}$.

\subsection{KKT Conditions of (P3)} \label{opt-A}
The Lagrangian function of (P3) is expressed as
\begin{align}
	&\mathcal{L} \left( \mbf{p}, \mbf{t}, \bsm{\lambda}, \bsm{\mu}, v, w \right) \nonumber \\
	=&-\sum_{s\in \mathcal{S}} \log\left(1+ m_s^{\sf r} p_s^{\sf r} t_s^2  \right) 
	- \sum_{i\in \mathcal{L}_3} \log\left(1+ m_{i}^{\sf d} p_{i}^{\sf d}  \right) \nonumber \\
	&-\sum_{s \in \mathcal{S}} \lambda_s^{\sf r} p_s^{\sf r} -\sum_{i \in \mathcal{L}_3} \lambda_i^{\sf d} p_i^{\sf d} 
	 + v \left(\sum_{s\in \mathcal{S}} p_s^{\sf r} + \sum_{i\in \mathcal{L}_3} p_i^{\sf d} - P\right) \nonumber \\
	& -\sum_{s\in \mathcal{S}} \mu_s t_s  + w \left(\sum_{s \in \mathcal{S}} t_s - 1\right),
\end{align}
where $\bsm{\lambda}^{\sf r} = [\lambda_1^{\sf r}, \cdots, \lambda_S^{\sf r}]^{T}$, $\bsm{\lambda}^{\sf d} = [\lambda_1^{\sf d}, \cdots, \lambda_{L_3}^{\sf d}]^{T}$, $\bsm{\mu} = [\mu_1, \cdots, \mu_S]^{T} $, $v$, and $w$ are the dual variables associated with constraints C8, C9, C11, C10, and C3, respectively, and $\bsm{\lambda} = [(\bsm{\lambda}^{\sf r})^T, (\bsm{\lambda}^{\sf d})^T]^T$.
The stationarity conditions are obtained by setting the first-order derivative of $\mathcal{L} \left( \mbf{p}, \mbf{t}, \bsm{\lambda}, \bsm{\mu}, v, w \right)$ w.r.t. $\mbf{p}$ and $\mbf{t}$ to zero, i.e.,
\begin{subnumcases}
	{}
	-\frac{ m_s^{\sf r} t_s^2}{1+ m_s^{\sf r} p_s^{\sf r} t_s^2} - \lambda_s^{\sf r} + v =0, ~~\forall s \in \mathcal{S}, \label{firstorder-1}\\
	-\frac{m_i^{\sf d} }{1+ m_{i}^{\sf d} p_{i}^{\sf d}} - \lambda_i^{\sf d} + v =0,~~\forall i \in \mathcal{L}_3, \label{firstorder-2}\\
	-\frac{2 m_s^{\sf r} p_s^{\sf r} t_s }{1+ m_s^{\sf r} p_s^{\sf r} t_s^2} - \mu_s + w =0, ~~\forall s \in \mathcal{S}. \label{firstorder-3}
\end{subnumcases}
The dual feasibility and complementary slackness conditions are given by
\begin{subnumcases}
	{}
	\lambda_s^{\sf r} \geq 0, ~~\forall s \in \mathcal{S}, \label{dual-0} \\
	\lambda_i^{\sf d} \geq 0,~~ \forall i \in \mathcal{L}_3,  \label{dual-1} \\
	\lambda_s^{\sf r} p_s^{\sf r} = 0, ~~\forall s \in \mathcal{S}, \label{complementary-0} \\
	\lambda_i^{\sf d} p_i^{\sf d} = 0, ~~\forall i \in \mathcal{L}_3, \label{complementary-1} \\
	\mu_s \geq 0, ~~\forall s \in \mathcal{S}, \label{dual-2} \\
	\mu_s t_s =0, ~~\forall s \in \mathcal{S}. \label{complementary-2}
\end{subnumcases}
The KKT solutions can be calculated via solving the conditions in \eqref{firstorder-1}-\eqref{firstorder-3} and \eqref{dual-0}-\eqref{complementary-2}, together with the primal feasibility given in the constraints of (P3).
As the KKT conditions are the first-order necessary condition for a solution to be optimal, we have the following properties on the optimal power and RIS partitioning strategy.
\begin{lemm}\label{lemm-linear-opt}
	The optimal $p_s^{\sf r}$ and the optimal $t_s$ to (P3) satisfy the following equation:
	\begin{align} \label{linear-relation}
		t_s = \frac{2v}{w} p_s^{\sf r} = \frac{1}{P^{\sf r}}p_s^{\sf r},~~\forall s \in \mathcal{S},
	\end{align}
	where $P^{\sf r} = \sum_{i \in \mathcal{S}} p_i^{\sf r}$ denotes the total power allocated to the cascaded channel.
\end{lemm}
\begin{IEEEproof}
	We first consider the case of $\left\{t_s>0, p_s^{\sf r}>0\right\}$.
	Combining \eqref{firstorder-1} and \eqref{firstorder-3} yields
	\begin{align} \label{linear-relation-derivation}
		t_s = \frac{2 \left(\lambda_s^{\sf r}-v\right)}{\mu_s-w}  p_s^{\sf r}.
	\end{align}
	From the complementary slackness conditions, we have $\lambda_s^{\sf r} = \mu_s = 0$.
	Then, summing up \eqref{linear-relation-derivation} for all $s \in \mathcal{S}$, we get $\frac{2v}{w} = \frac{1}{P^{\sf r}}$,
	and thus \eqref{linear-relation} is obtained.
	Second, \eqref{linear-relation} holds trivially if $p_s^{\sf r} = t_s = 0$.
	It remains to consider the cases of $\{t_s = 0, p_s^{\sf r} \neq 0\}$ and $\{t_s \neq 0, p_s^{\sf r} = 0\}$.
	These two cases are impossible for an optimal pair of $\{t_s, p_s^{\sf r}\}$ since they
	cause a waste of either power or reflecting resource.
	This completes the proof.
\end{IEEEproof}

\begin{lemm}\label{corol-descend}
	Suppose $m_1^{\sf r} \geq m_2^{\sf r} \geq \cdots \geq m_S^{\sf r}$.
	Then, the entries of the optimal RIS partitioning $\mbf{t}$ and the optimal power allocation $\mbf{p}^{\sf r}$ are both arranged in the descending order, i.e., $t_1 \geq t_2 \geq \cdots \geq t_S$ and $p_1^{\sf r} \geq p_2^{\sf r} \geq \cdots \geq p_S^{\sf r}$.
\end{lemm}
\begin{IEEEproof}
	Suppose without loss of generality that $p_i^{\sf r} < p_j^{\sf r}$ for some $j > i$, where $i,j \in \mathcal{S}$.
	From Lemma \ref{lemm-linear-opt}, we have $t_i < t_j$.
	For the case of $m_i^{\sf r} = m_j^{\sf r}$, $p_i^{\sf r} < p_j^{\sf r}$ and $t_i < t_j$ are obviously not the optimal solution.
	For the case of $m_i^{\sf r} > m_j^{\sf r}$, denote the optimal power allocation and RIS partitioning as $\mbf{p}$ and $\mbf{t}$, respectively.
	Let $\hat{\mbf{p}} = [p_1^{\sf r}, \cdots, p_{i-1}^{\sf r},  p_j^{\sf r},  p_{i+1}^{\sf r}, $ $\cdots,$ $ p_{j-1}^{\sf r}, p_i^{\sf r}, p_{j+1}^{\sf r},  \cdots, p_S^{\sf r}, p_1^{\sf d}, \cdots, p_{L_3}^{\sf d} ]^T$ denote a modified transmit power vector, where the only difference between the optimal $\mbf{p}$ lies in the exchanged ordering of $p_i^{\sf r}$ and $p_j^{\sf r}$.
	Similarly, we construct a modified RIS partitioning strategy as $\hat{\mbf{t}} = [t_1, \cdots, t_{i-1},  t_j,  t_{i+1}, \cdots, t_{j-1}, t_i, t_{j+1},  \cdots, t_S]^{T}$.
	The difference between $C\left(\hat{\mbf{p}}, \hat{\mbf{t}} \right)$ and $C\left(\mbf{p}, \mbf{t} \right)$ is given by
	\begin{align}
		&C\left(\hat{\mbf{p}}, \hat{\mbf{t}} \right) - C\left(\mbf{p}, \mbf{t} \right) \nonumber \\
		=&  \log\left(1+ \frac{\left(m_i^{\sf r}-m_j^{\sf r}\right) \left( p_j^{\sf r} t_j^{2} - p_i^{\sf r} t_i^{2}  \right) }{ \left(1+ m_i^{\sf r} p_i^{\sf r} t_i^{2}  \right) \left(1+ m_j^{\sf r} p_j^{\sf r} t_j^{2} \right) }  \right) > 0,
	\end{align}
	which leads to a contradiction.
\end{IEEEproof}

We next establish the optimal solution to (P3) based on Lemmas \ref{lemm-linear-opt} and \ref{corol-descend}.
To begin with, we consider two sub-problems of (P3) in the following.

\subsection{Optimal Solutions to  (P3.1) and (P3.2)}
We construct two sub-problems (P3.1) and (P3.2). 
Specifically, (P3.1) is to optimize $\mbf{p}$ with fixed $\mbf{t}$, i.e.,
\begin{align}
	\textrm{(P3.1):} ~~\max_{ \mbf{p} } \quad &  C\left(\mbf{p}\right)  \nonumber \\ 
	\operatorname{ s.t. } \quad
	& \textrm{C8}, \textrm{C9}, \textrm{C10} ,\nonumber
\end{align}
where $C(\mbf{p})$ is an abbreviation of $C(\mbf{p}, \mbf{t})$ for fixed $\mbf{t}$.
The KKT conditions of (P3.1) are given by \eqref{firstorder-1}, \eqref{firstorder-2}, \eqref{dual-0}, \eqref{dual-1}, \eqref{complementary-0}, \eqref{complementary-1}, C8, C9, and C10.
The other sub-problem (P3.2) is constructed as
\begin{align}
	\textrm{(P3.2):} ~~ \max_{ \mbf{t} } \quad &  C\left(\mbf{t}\right) \triangleq \sum_{s\in \mathcal{S}} \log\left(1+ \widetilde{m}_s t_s^2  \right) \nonumber \\ 
	\operatorname{ s.t. } \quad
	& \textrm{C11}, \textrm{C3}, \nonumber
\end{align}
where $\widetilde{m}_s \triangleq m_s p_s^{\sf r}$ and $C(\mbf{t})$ is given by $C(\mbf{p}, \mbf{t})$ with $\mbf{p}$ fixed and the constant terms omitted.
From Lemma \ref{corol-descend}, we assume without loss of generality that $\{\widetilde{m}\}_{s=1}^S$ are arranged in the descending order.
The KKT conditions of (P3.2) are given by \eqref{firstorder-3}, \eqref{dual-2}, \eqref{complementary-2}, C11, and C3.
Clearly, the KKT conditions of (P3.1) and (P3.2) are two complementary subsets of those of (P3).

\newcounter{TempEqCnt} 
\setcounter{TempEqCnt}{\value{equation}} 
\setcounter{equation}{32}
\begin{figure*}[b]
	\hrulefill
	\begin{equation} \label{opt-t-2}
		\mbf{t}=
		\begin{cases}
			[1,0]^T, & 2+\sqrt{1-\frac{\widetilde{m}_2}{\widetilde{m}_1}} \geq\sqrt{\widetilde{m}_2} ,\\
			\arg \max_{\mbf{t}} \left\{ C \left([1,0]^T \right) , C \left( \mbf{t}^{\left\{+,+\right\}} \right) \right\}, & 2+\sqrt{1-\frac{\widetilde{m}_2}{\widetilde{m}_1}} < \sqrt{\widetilde{m}_2}.
		\end{cases} \\
	\end{equation}
\end{figure*}
\setcounter{equation}{\value{TempEqCnt}}

\subsubsection{Optimal Solution to  (P3.1)}
 (P3.1) is convex and its optimal solution is readily given by the {\textit{water-filling strategy}} \cite{Goldsmith2005Wireless}:
\begin{align} \label{water-filling-r}
	p_s^{\sf r} =
	\begin{cases}
		0,& \lambda_s^{\sf r} > 0,\\
		\frac{1}{v}-\frac{1}{m_s^{\sf r} t_s^{2}}, &\lambda_s^{\sf r} = 0,
	\end{cases}
\end{align}
and
\begin{align} \label{water-filling-d}
	p_i^{\sf d} =
	\begin{cases}
		0,& \lambda_i^{\sf d} > 0,\\
		\frac{1}{v}-\frac{1}{m_i^{\sf d}}, &\lambda_i^{\sf d} = 0.
	\end{cases}
\end{align}

\subsubsection{Optimal Solution to  (P3.2)}
(P3.2) is non-convex since the objective function is non-concave w.r.t. $\mbf{t}$.
Here, we obtain the global maximum of (P3.2) by carefully analysing the KKT conditions. 
\begin{lemm}\label{kkt-points}
	The optimal solution to (P3.2) takes the form of
	\begin{equation} \label{KKT-patterns}
		t_s=
		\begin{cases}
			0,& \mu_s > 0,\\
			\frac{1}{w}  \pm \sqrt{\frac{1}{w^{2}} - \frac{1}{\widetilde{m}_s}} , &\mu_s = 0.
		\end{cases} \\
	\end{equation}
\end{lemm}
\begin{IEEEproof}
	The optimal solution to (P3.2) satisfies the KKT conditions \eqref{firstorder-3}, \eqref{dual-2} and \eqref{complementary-2}.
	If $\mu_s >0$, we obtain $t_s = 0$ from \eqref{complementary-2}.
	Otherwise, $\mu_s = 0$ implies $t_s \geq 0$. 
	The discussion of $\mu_s = 0$ is thus divided into the cases of $t_s > 0$ and $t_s = 0$. 
	If $t_s>0$, we have $w = \frac{2 \widetilde{m}_s t_s}{1+\widetilde{m}_s t_s^2} >0$ by \eqref{firstorder-3}, and correspondingly $t_s = \frac{1}{w}  \pm \sqrt{\frac{1}{w^{2}} - \frac{1}{\widetilde{m}_s}}$.
	If $t_s=0$, we obtain $w = \frac{2\widetilde{m}_s t_s}{1+\widetilde{m}_s t_s^2} =0$.
	However, there always exists a $t_j>0$ ($j \neq s$) such that $\mu_j = \frac{2\widetilde{m}_j t_j}{1+\widetilde{m}_j t_j^2} > 0$, which contradicts the complementary slackness condition by noting $\mu_j t_j >0$.
	To summarize, the KKT conditions of  (P3.2) are satisfied with $t_s = 0$ if $\mu_s>0$, and with $t_s = \frac{1}{w}  \pm \sqrt{\frac{1}{w^{2}} - \frac{1}{\widetilde{m}_s}}$ if $\mu_s=0$.
	This completes the proof.
\end{IEEEproof}

For notation simplicity, we abbreviate the three possible expressions of each $t_s$ in Lemma \ref{kkt-points} as $t_s^0 \triangleq 0$, $t_s^+ \triangleq \frac{1}{w}  + \sqrt{\frac{1}{w^{2}} - \frac{1}{\widetilde{m}_s}}$, and $t_s^- \triangleq \frac{1}{w}  - \sqrt{\frac{1}{w^{2}} - \frac{1}{\widetilde{m}_s}}$.
Define $\mbf{t}^{\left\{\mathsf{l}_1,\mathsf{l}_2, \cdots, \mathsf{l}_S\right\}} \triangleq [t_1^{\mathsf{l}_1}, t_2^{\mathsf{l}_2}, \cdots, t_S^{\mathsf{l}_S} ]^T$ as a candidate solution to (P3.2), where $\mathsf{l}_1,\mathsf{l}_2, \cdots, \mathsf{l}_S \in \left\{0,+,-\right\}$.
Note that for given $\{\widetilde{m}_s\}_{s=1}^S$, 
a specific pattern $\mbf{t}^{\left\{\mathsf{l}_1,\mathsf{l}_2, \cdots, \mathsf{l}_S\right\}}$ is not necessarily a valid solution to (P3.2).
We need to verify the validity of $\mbf{t}^{\left\{\mathsf{l}_1,\mathsf{l}_2, \cdots, \mathsf{l}_S\right\}}$ by checking the feasibility of C3.
Also, since C3 possibly has more than one solution, each pattern $\mbf{t}^{\left\{\mathsf{l}_1,\mathsf{l}_2, \cdots, \mathsf{l}_S\right\}}$ may correspond to multiple solutions.

We now show that there is no need to exhaustively search over all the possible $3^S$ patterns specified by \eqref{KKT-patterns}.
In fact, the number of valid solutions is very limited, which allows us to readily find the global optimum.
We first consider the special case that the RIS is partitioned into two sub-surfaces.

\begin{lemm}\label{opt-2}
	For $S=2$, the optimal solution to (P3.2) occurs only at $[1,0]^T$ and $\mbf{t}^{\left\{+,+\right\}}$,
	which is given by \eqref{opt-t-2} at the bottom of this page.
\end{lemm}
\begin{IEEEproof}
	Please refer to Appendix B.
\end{IEEEproof}
Proposition \ref{opt-p3b} generalizes the results in Lemma \ref{opt-2} to arbitrary $S$ sub-surfaces.

\begin{prop} \label{opt-p3b}
For an arbitrary number $S$ of sub-surfaces, the optimal solution to (P3.2) can appear only at $[1,0,0,\cdots,0]^T$, $\mbf{t}^{\left\{+, +, 0, \cdots, 0\right\}}$, $\mbf{t}^{\left\{+, +,+, \cdots, 0\right\}}$, $\cdots$, $\mbf{t}^{\left\{+, +, +,\cdots, +\right\} }$, i.e.,
the optimal solution is given by
\setcounter{TempEqCnt}{\value{equation}} 
\setcounter{equation}{33}
\begin{align} \label{eq-prop1}
	\mbf{t} &= \arg \max_{\mbf{t}} \Big\{ C \left( [1,0,0,\cdots,0]^T\right) , C \left( \mbf{t}^{\left\{+, +, 0, \cdots, 0\right\} }\right), \nonumber \\
	& ~~~~C \left( \mbf{t}^{\left\{+, +, +,\cdots, 0\right\} }\right), \cdots,  C \left( \mbf{t}^{\left\{+, +, +,\cdots, +\right\} }\right) \Big\}.
\end{align}
\end{prop}
\begin{IEEEproof}
	Please refer to Appendix C.
\end{IEEEproof}

\begin{figure*}[t]
	\vspace{-.5em}
	\centering
	\subfigure[]
	{\includegraphics[width=.9\columnwidth]{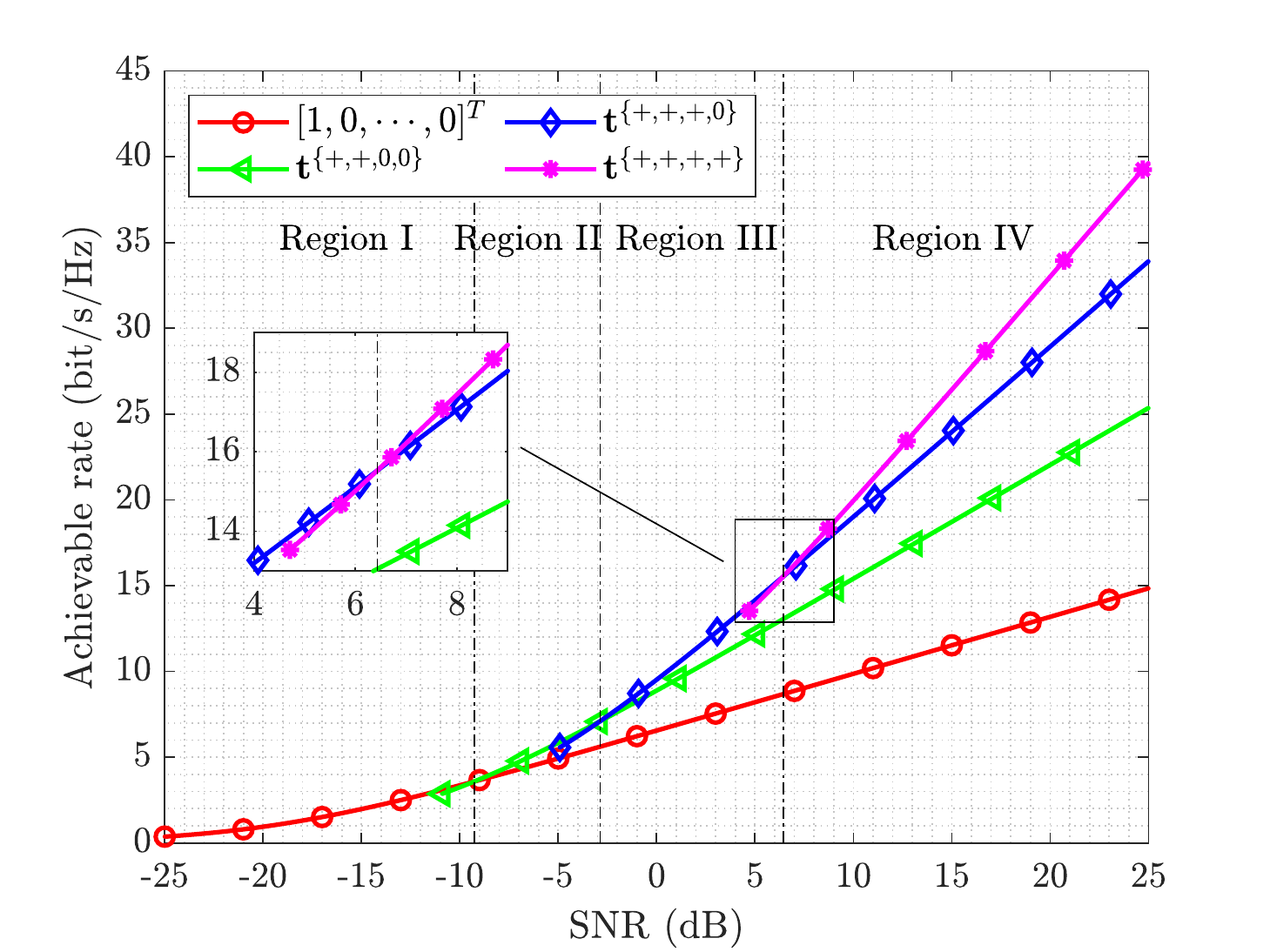}\label{fig-Prop1b}} \hfil
	\subfigure[]
	{\includegraphics[width=.9\columnwidth]{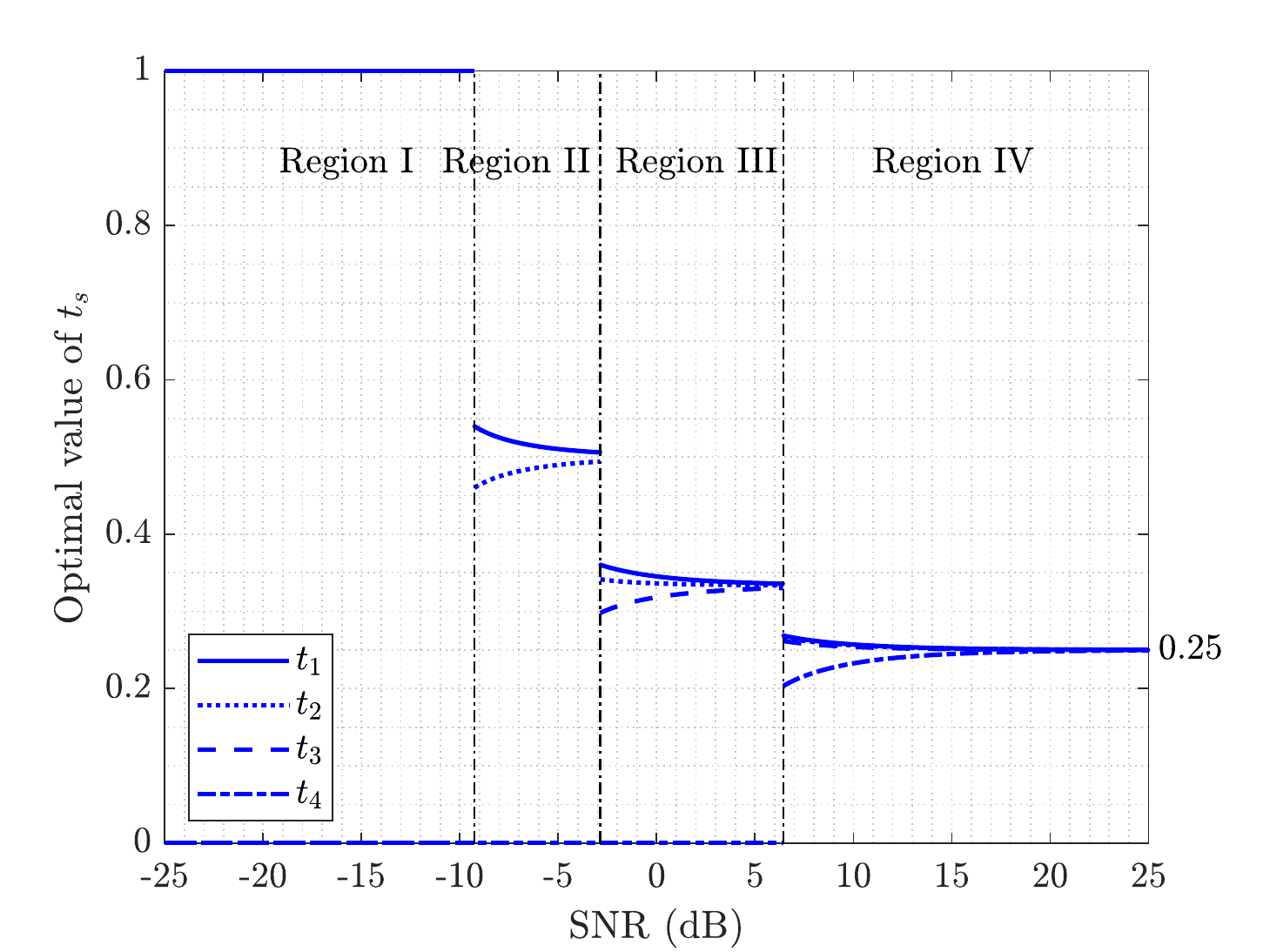}\label{fig-Prop1d}}
	\caption{Numerical illustrations to Proposition \ref{opt-p3b}, where $\min \{L_1, L_2\} = 4$, $m_1 = 93$, $m_2 = 74$, $m_3 = 54$, and $m_4 = 15$.}
	\label{fig-Prop1}
\end{figure*}

To better understand Proposition \ref{opt-p3b}, in Fig. \ref{fig-Prop1}, we plot the achievable rate and the corresponding optimal solution over different transmit SNR $\frac{P}{\sigma^2}$.
The coefficients of the paired Tx-RIS-Rx paths are set as $m_1 = 93$, $m_2 = 74$, $m_3 = 54$, and $m_4 = 15$.
It is assumed that the Tx-Rx direct channel is totally blocked, and the transmit power is equally allocated to the four paired paths.
In Regions I, II, III, and IV, the optimal solution is given by activating the first one, two, three, and four paired paths, respectively. 
In Fig. \ref{fig-Prop1b}, we show that for a given SNR, not every pattern provides a valid solution.
It is also observed that multiple solutions (corresponding to multiple patterns) may exist at a given SNR. 
For instance, the pattern $\mbf{t}^{\{+,+,+,+\}}$ appears when the SNR is greater than $4.71$ dB, but it becomes the optimal solution only when the SNR exceeds $6.43$ dB.
This explains why the comparison of multiple patterns is required in \eqref{eq-prop1} to find the optimum.
Moreover, we see in Fig. \ref{fig-Prop1d} that at a relatively low SNR, to maximize the achievable rate, the cascaded channel prefers to concentrate the reflecting resources on a small number of the paired paths, which behaves similar to the water-filling solution.
Similarly, when the SNR is high, the cascaded channel prefers to evenly assign the reflecting resources to all the paired paths.

\subsection{Optimal Solution to (P3)}
The optimal solution to (P3) is necessarily the optimal solutions to (P3.1) and (P3.2).
Besides, Lemma \ref{lemm-linear-opt} also provides a necessary condition for a solution to be optimal.
In the following, we combine these necessary conditions to describe the optimal solution to (P3).
From Proposition \ref{opt-p3b}, the search over a different number of sub-surfaces, or equivalently, a different number of activated Tx-RIS-Rx paths, is required to find the optimum to (P3.2).
For this reason, it is necessary to search the optimal solution to (P3) over a different number of activated Tx-RIS-Rx paths and Tx-Rx paths.
Denote by $\mathcal{S}_{\sf a}$ the index set of the activated Tx-RIS-Rx paths, and by $\mathcal{I}_{\sf a}$ that of the activated Tx-Rx paths.
We have $\mathcal{S}_{\sf a} \in \mathcal{S}_{\sf collection} \triangleq \left\{ \{1\}, \{1,2\},\cdots,\{1,2,\cdots,\min \{L_1, L_2\}\} \right\}$ and $\mathcal{I}_{\sf a} \in \mathcal{I}_{\sf collection} \triangleq \left\{\{1\}, \{1,2\},\cdots,\{1,2,\cdots,L_3\} \right\}$.\footnote{
	For the case of $\mathcal{S}_{\sf a} = \{1\}$, i.e., $\mbf{t} = [1,0,0,\cdots,0]^T$, the corresponding power allocation can be readily obtained by the water-filling strategy in \eqref{water-filling-r} and \eqref{water-filling-d}, and there is no need to run the subsequently described procedures.
}
Given $\mathcal{S}_{\sf a}$ and $\mathcal{I}_{\sf a}$, combining the aforementioned necessary conditions yields the following system of nonlinear equations:
\begin{subnumcases}
	{\label{nonlinear-equations} }
	p_s^{\sf r} = \frac{1}{v}-\frac{1}{m_s^{\sf r} t_s^{2}},~~ s \in \mathcal{S}_{\sf a}, \label{lm-method-1} \\
	p_i^{\sf d} = \frac{1}{v}-\frac{1}{m_i^{\sf d}}, ~~i \in \mathcal{I}_{\sf a}, \label{nonlinear-equations-pi} \\
	t_s = \frac{1}{w}  + \sqrt{\frac{1}{w^{2}} - \frac{1}{m_s^{\sf r} p_s^{\sf r}}},~~ s \in \mathcal{S}_{\sf a}, \label{nonlinear-equations-ts}\\
	t_s = \frac{2v}{w} p_s^{\sf r} = \frac{1}{P^{\sf r}}p_s^{\sf r}, ~~s \in \mathcal{S}_{\sf a}, \label{nonlinear-equations-linear-relation} \\
	\sum_{s\in \mathcal{S}_{\sf a}} p_s^{\sf r} + \sum_{i\in \mathcal{I}_{\sf a}} p_i^{\sf d} = P, \label{nonlinear-equations-power-budget} \\
	\sum_{s\in \mathcal{S}_{\sf a}} t_s = 1.
\end{subnumcases}
The optimal solution to (P3) necessarily satisfies \eqref{nonlinear-equations}.
By calculating and comparing all the feasible solutions to \eqref{nonlinear-equations}, we can reach the optimal $\mbf{p}$ and $\mbf{t}$ of (P3).

In the following, we propose a 1D grid search algorithm that outputs an approximately optimal solution to (P3).
A key observation of \eqref{nonlinear-equations} is that, for given $v$, we immediately obtain the power allocated to each Tx-Rx path from \eqref{nonlinear-equations-pi}, and the total power allocated to the cascaded channel is thus obtained according to \eqref{nonlinear-equations-power-budget}.
Moreover, substituting \eqref{nonlinear-equations-linear-relation} into \eqref{nonlinear-equations-ts} yields the following cubic equation for each $s$:
\begin{align} \label{cubic}
	\left(p_s^{\sf r}\right)^3 - \frac{1}{v} \left(p_s^{\sf r}\right)^2 +\frac{(P^{\sf r})^2}{m_s^{\sf r}} = 0,~~s \in \mathcal{S}_{\sf a},
\end{align}
where $p_s^{\sf r}$ is a valid solution only when $p_s^{\sf r} \geq \max \left\{\frac{4v^2(P^{\sf r})^2}{m_s^{\sf r}}, \frac{1}{2v} \right\}$, by considering that the term in the square root of \eqref{nonlinear-equations-ts} is non-negative. 
Solve \eqref{cubic} for all $s$ and then substitute all the valid solutions into \eqref{nonlinear-equations-power-budget} to verify whether the total power budget is satisfied.
If satisfied, we calculate $\mbf{t}$ and $w$ by \eqref{nonlinear-equations-linear-relation} with the obtained $\mbf{p}$.
In the above process, 
a 1D grid search over $v$ is required to find all the possible solutions to \eqref{nonlinear-equations}.
We provide the implementation details in Algorithm \ref{alg1}.
In Line \ref{lb}, the lower bound of $v$ in the search (denoted as $b_{\sf l}$) is given by the maximum of the two arguments, where the first is obtained by plugging \eqref{nonlinear-equations-pi} into the inequation $\sum_{s \in \mathcal{S}_{\sf a}} p_s^{\sf r} \leq P$, and the second is obtained since $P \geq p_s^{\sf r} \geq \frac{1}{2v}$.
In Line \ref{ub}, the upper bound of $v$ in the search (denoted as $b_{\sf u}$) is given by the minimum of the two arguments, where the first is given by applying $t_s \leq 1$ to \eqref{lm-method-1}, \eqref{nonlinear-equations-pi}, and \eqref{nonlinear-equations-power-budget}, and the second is given by applying $p_i^{\sf d} \geq 0$ to \eqref{nonlinear-equations-pi}.
From Line 8 to 14, the algorithm traverses $v$s from the search lower bound $b_{\sf l}$ to the search upper bound $b_{\sf u}$ with the grid size $s_{\sf grid}$.

\begin{algorithm}[t]\label{alg1}
	\SetAlgoLined
	\SetKw{KwInput}{Input:}
	\SetKw{KwOutput}{Output:}
	\KwInput{
		$\left\{m_s^{\sf r}\right\}_{s=1}^{\min \{L_1, L_2\}}$, $\left\{m_i^{\sf d}\right\}_{i=1}^{L_3}$, $P$, grid size $s_{\sf grid}$, accuracy tolerance $\epsilon_{\sf acc}$;\
	}
	
	\KwOutput{
		Optimal solution pair $(\mbf{p}, \mbf{t})$\;
	}
	
	Calculate $\mbf{p}$ with $\mbf{t} = [1,0,0,\cdots,0]^T$ by \eqref{water-filling-r} and \eqref{water-filling-d}, and then save $(\mbf{p}, \mbf{t})$;
	
	\For{$\mathcal{S}_{\sf a} \in \mathcal{S}_{\sf collection}/\{1\} $}
	{\For{$\mathcal{I}_{\sf a} \in \mathcal{I}_{\sf collection} $}
		{
			Calculate the search lower bound $b_{\sf l} = \max \left\{\frac{|\mathcal{I}_{\sf a}|}{P + \sum_{i\in \mathcal{I}_{\sf a}} \frac{1}{m_i^{\sf d}}}, \frac{1}{2P}\right\}$; \label{lb} \
			
			Calculate the search upper bound $b_{\sf u} = \min \left\{\frac{|\mathcal{I}_{\sf a}| + |\mathcal{S}_{\sf a}| }{P+\sum_{i\in \mathcal{I}_{\sf a}} \frac{1}{m_i^{\sf d}} + \sum_{s\in \mathcal{S}_{\sf a}} \frac{1}{m_s^{\sf r}}}, m_{|\mathcal{I}_{\sf a}|}^{\sf d}\right\}$; \label{ub}\
			
			\For{$v = b_{\sf l}: s_{\sf grid}: b_{\sf u}$}
			{ 	Calculate $p_i^{\sf d}$ by \eqref{nonlinear-equations-pi}, $\forall i \in \mathcal{I}_{\sf a}$; \
				
				Solve the cubic equation \eqref{cubic}, $\forall s \in \mathcal{S}_{\sf a}$; \
				
				Collect real $p_s^{\sf r}$ that satisfies $p_s^{\sf r} \geq \max \left\{\frac{4v^2(P^{\sf r})^2}{m_s^{\sf r}}, \frac{1}{2v} \right\}$, $\forall s \in \mathcal{S}_{\sf a}$; \
				
				Collect all the feasible $\mbf{p}$ that satisfies $\frac{\left| \sum_{s\in \mathcal{S}_{\sf a}} p_s^{\sf r} + \sum_{i\in \mathcal{I}_{\sf a}} p_i^{\sf d} - P \right|}{P} < \epsilon_{\sf acc}$; \
				
				Calculate $\mbf{t}$ for each feasible $\mbf{p}$ by \eqref{nonlinear-equations-linear-relation}, and then save each $(\mbf{p}, \mbf{t})$; \								
			}
		}
	}
	Return $(\mbf{p}, \mbf{t})$ which yields the maximum achievable rate.
	\caption{Proposed 1D Grid Search Algorithm}
\end{algorithm}

The computational complexity of the 1D grid search algorithm is analyzed as follows.
At each iteration, the operations in Line 9 and Line 10 have a complexity of $\mathcal{O}\left(|\mathcal{I}_{\sf a}|\right)$ and $\mathcal{O}\left(|\mathcal{S}_{\sf a}|\right)$, respectively.
According to the Vieta's formulas, the cubic equation \eqref{cubic} has one negative solution.
This indicates that for each $s \in \mathcal{S}_{\sf a}$, the number of solutions satisfying the condition in Line 11 is at most two.
Consequently, the number of $\mbf{p}$s needed to be verified in Line 12 is at most $2^{|\mathcal{S}_{\sf a}|}$.
This corresponds to the worst-case complexity of Algorithm \ref{alg1}, given as $\mathcal{O}\left( \lfloor \frac{b_{\sf u} - b_{\sf l}}{s_{\sf grid}} \rfloor \sum_{\mathcal{I}_{\sf a} \in \mathcal{I}_{\sf collection} } \sum_{\mathcal{S}_{\sf a} \in \mathcal{S}_{\sf collection}/\{1\} } \left(|\mathcal{I}_{\sf a}| + |\mathcal{S}_{\sf a}| + 2^{|\mathcal{S}_{\sf a}|} \right) \right)$.
By calculating the summation and ignoring the lower-order terms,
the worst-case complexity can be simplified to
$\mathcal{O}\big(\lfloor \frac{b_{\sf u} - b_{\sf l}}{s_{\sf grid}} \rfloor ( L_3^2 \min \{L_1, L_2\} +  2^{\min \{L_1, L_2\}} L_3 ) \big)$.
We note that the above worst-case analysis overestimates the actual complexity.
In our simulation, the cubic equation \eqref{cubic} usually has only one or no solution satisfying the condition in Line 11.
This corresponds to the best case where at most a single $\mbf{p}$ is processed in Line 12.
The best-case complexity of Algorithm \ref{alg1} is given by $\mathcal{O}\big(\lfloor \frac{b_{\sf u} - b_{\sf l}}{s_{\sf grid}} \rfloor (L_3^2 \min \{L_1, L_2\} + \min \{L_1, L_2\}^2 L_3) \big)$, which is in accordance with our empirical experience of running the algorithm in practice.

As a low-complexity alternative, the LM method \cite{more1978levenberg}, a trust region approach that synthesizes the steepest descent and Gaussian-Newton methods can be adopted to find a solution to \eqref{nonlinear-equations}, which is also a stationary point to (P3).
The computational complexity of the LM method is given by $\mathcal{O} \left(\min \{L_1, L_2\} L_3 \epsilon^{-2}\right)$, where $\epsilon$ denotes the given convergence accuracy \cite{LM_complexity}.
Simulation results show that the performance of the LM method is close to that of Algorithm \ref{alg1}.

\subsection{Optimal Path Pairing Strategy} \label{opt-pairing}
Proposition \ref{opt-path-pairing} gives the optimal path pairing matrix $\mbf{B}$ in closed form.
\begin{prop} \label{opt-path-pairing}
	Assume without loss of generality that $L_1 \leq L_2$.
	The optimal path pairing matrix is given by $\mbf{B} = \left[\mbf{I}_{L_1 \times L_1}, \bsm{0} \right]$. 
\end{prop}
\begin{IEEEproof}
	The result is trivial when $L_1=1$.
	When $L_1=2$, we only need to compare two different configurations of the pairing matrix, i.e.,
	\begin{align} \label{B2-path-pairing-1}
		\mbf{B} = \begin{bmatrix}
			1 & 0 & 0 & \cdots & 0  \\
			0 & 1 & 0 & \cdots & 0  \\
		\end{bmatrix},
	\end{align}
	and
	\begin{align} \label{B2-path-pairing-2}
		\check{\mbf{B}} = \begin{bmatrix}
			0 & 1 & 0 & \cdots & 0  \\
			1 & 0 & 0 & \cdots & 0  \\
		\end{bmatrix}.
	\end{align}
	Applying $\mbf{B}$ results in $m_s^{\sf r} = c\left|\alpha_s \beta_s\right|^2$, $s = 1,2$, and applying $\check{\mbf{B}}$ results in $m_1^{\sf r} = c\left|\alpha_1 \beta_2\right|^2$ and $m_2^{\sf r} = c\left|\alpha_2 \beta_1\right|^2$,
	where $c = \mathrm{PL}^{\sf r}\frac{ M_{\sf t} M_{\sf r} N^2 }{L_1 L_2\sigma^2}$.
	Denote the optimal power and RIS partitioning associated with $\check{\mbf{B}}$ as $\check{\mbf{p}} = [\check{p}_1^{\sf r}, \check{p}_2^{\sf r},\check{p}_1^{\sf d},\cdots, \check{p}_{L_3}^{\sf d}]^T$ and $\check{\mbf{t}} = [\check{t}_1, \check{t}_2]^T$, respectively.
	We have
	\begin{align} \label{eq-prop2-proof}
		&C\left(\check{\mbf{p}}, \check{\mbf{t}}, \mbf{B}\right) - C\left(\check{\mbf{p}}, \check{\mbf{t}}, \check{\mbf{B}}\right) \nonumber \\
		=& \log \left(1+ \frac{c \left(\left|\beta_1\right|^2-\left|\beta_2\right|^2\right) \left(\left|\alpha_1\right|^2 \check{p}_1^{\sf r} \check{t}_1^2-\left|\alpha_2\right|^2 \check{p}_2^{\sf r} \check{t}_2^2\right)}{\left(1+c \left|\alpha_1 \beta_2\right|^2 \check{p}_1^{\sf r} \check{t}_1^2\right) \left(1+c \left|\alpha_2 \beta_1\right|^2 \check{p}_2^{\sf r} \check{t}_2^2\right) } \right).
	\end{align} 
	Suppose without loss of generality that $\left|\alpha_1 \beta_2\right| \geq \left|\alpha_2 \beta_1\right|$, we have $\check{p}_1^{\sf r} \geq \check{p}_2^{\sf r}$ and $\check{t}_1 \geq \check{t}_2$ according to Lemma \ref{corol-descend}.
	Recall that $\left|\alpha_1\right| \geq \left|\alpha_2\right|$ and $\left|\beta_1\right| \geq \left|\beta_2\right|$, we obtain $C\left(\check{\mbf{p}}, \check{\mbf{t}}, \mbf{B}\right) - C\left(\check{\mbf{p}}, \check{\mbf{t}}, \check{\mbf{B}}\right)\geq 0$ from \eqref{eq-prop2-proof}.
	Moreover, denote by $\mbf{p}$ and $\mbf{t}$ the optimal power and RIS partitioning associated with $\mbf{B}$, respectively.
	We have $C\left(\mbf{p}, \mbf{t}, \mbf{B}\right) \geq C\left(\check{\mbf{p}}, \check{\mbf{t}}, \mbf{B}\right) \geq C\left(\check{\mbf{p}}, \check{\mbf{t}}, \check{\mbf{B}}\right)$, which completes the proof of the case $L_1=2$.

	When $L_1 >2$, suppose that the path pairing matrix is not chosen as in Proposition \ref{opt-path-pairing}.
	Then, there always exist two Tx-RIS-Rx path pairs with the corresponding path pairing submatrix given by $\check{\mbf{B}}$ in \eqref{B2-path-pairing-2}.
	By following the argument for the case of $L_1 = 2$, we can replace the submatrix to $\mbf{B}$ in \eqref{B2-path-pairing-1} which do not decrease the achievable rate.
	Finally, we arrive at the path paring matrix given in Proposition \ref{opt-path-pairing}.
	This concludes the proof.
\end{IEEEproof}

Since $\left\{\alpha_\ell\right\}_{\ell=1}^{L_1}$ and $\left\{\beta_\ell\right\}_{\ell=1}^{L_2}$ are both arranged in the descending order of their magnitudes,
Proposition \ref{opt-path-pairing} shows that the Tx-RIS and RIS-Rx paths are paired in the same order based on their respective path gains.\footnote{The derived path pairing strategy resembles the subcarrier pairing results in MIMO relay systems. Interested readers can refer to \cite{MIMOrelay1, MIMOrelay2} for more details.}

\subsection{Algorithm Summary and Discussion}
For clarity, we summarize the overall algorithm for solving (P2) as follows.
The optimal $\mbf{B}$ of (P2) can be obtained by Proposition \ref{opt-path-pairing} in closed form regardless of the values of $\mbf{t}$ and $\mbf{p}$.
With the optimal $\mbf{B}$ readily at hand, we only need to solve (P3), whose optimal solution is
defined by one of the solutions to \eqref{nonlinear-equations} which yields the largest data rate.
We proposed two algorithms to efficiently solve \eqref{nonlinear-equations}, namely the 1D grid search algorithm which outputs an approximately optimal solution, and the LM method which reaches a stationary point.

\begin{table*}[t]
	\caption{Computational Complexity Comparisons}
	\label{computational-complexity}
	\centering
		\begin{tabular}{||c|c|c|c||}
			\hline
			\multicolumn{3}{||c|}{Algorithm} & Computational complexity \\ 
			\hline
			\multirow{3}{*}{\makecell{RIS partitioning \\ based methods}} & \multirow{2}{*}{1D search}  & Best case & $\mathcal{O}\left(\lfloor \frac{b_{\sf u} - b_{\sf l}}{s_{\sf grid}} \rfloor \left(L_3^2 \min \{L_1, L_2\} + \min \{L_1, L_2\}^2 L_3\right) \right)$  \\
			\cline{3-4}
			& & Worst case& $\mathcal{O}\left(\lfloor \frac{b_{\sf u} - b_{\sf l}}{s_{\sf grid}} \rfloor \big( L_3^2 \min \{L_1, L_2\} + 2^{\min \{L_1, L_2\}} L_3 \big) \right)$ \\
			\cline{2-4}
			& \multicolumn{2}{c|}{LM method} & $\mathcal{O} \Big(\min \{L_1, L_2\} L_3 \epsilon^{-2}\Big)$  \\
			\hline
			\multirow{2}{*}{\makecell{Element-wise \\ optimization methods}} &\multicolumn{2}{c|}{PB-element-wise AO \cite{ShuowenZhang2020JSAC}} & $\mathcal{O}\Big(I_0 \left(\left(3 M_{\sf r}^3  + 2 M_{\sf r}^2 M_{\sf t} + M_{\sf t}^2\right)N + M_{\sf r} M_{\sf t} \min(M_{\sf r}, M_{\sf t}) \right) \Big)$ \\
			\cline{2-4}
			& \multicolumn{2}{c|}{PB-WMMSE \cite{CunhuaPan2020TWC}} & $\mathcal{O}\Big(I_1 \left(M_{\sf t}^3 + M_{\sf t}^2 M_{\sf r} + N^3 + I_2 N^2 \right)\Big)$  \\
			\hline
		\end{tabular}
\end{table*}

\subsubsection{Algorithm Design in Finite-Size Systems}
We now discuss how to apply the optimization result of (P2) to the finite-size system design in (P1).
Firstly, $t_s N_{\sf y}$ may be fractional, and hence is not a feasible solution to (P1). 
We apply the rounding technique to transform each $t_s N_{\sf y}$ into an integer.
Secondly, the optimization of common phase shifts $\bsm{\psi}$ is not considered in (P2), as an arbitrary value of $\bsm{\psi}$ in the feasible region is asymptotically optimal.
However, a different value of $\bsm{\psi}$ results in a different achievable rate in (P1) for a finite-size system.
The problem of optimizing $\bsm{\psi}$ can be solved by existing techniques, e.g., the weighted minimum mean-square error (WMMSE) method \cite{shi-mmse}.
Different from the work \cite{CunhuaPan2020TWC} that utilizing the WMMSE method to handle the high-dimensional RIS reflection coefficients directly, we only need to solve $\bsm{\psi}$ in $S_{\min}^\star \leq \min \{L_1, L_2\}$ dimensions with much lower complexity.
We also note that optimizing $\bsm{\psi}$ becomes less critical when the system size is relatively large.
In fact, setting $\bsm{\psi}$ arbitrarily only brings negligible performance loss in the RIS-aided large-scale MIMO system. 
Thus, we choose $\bsm{\psi}$ randomly as our default setting in the computational complexity comparison part.

\subsubsection{Computational Complexity Comparisons}

In Table \ref{computational-complexity}, we compare the computational complexity of the proposed design with two element-wise optimization methods \cite{ShuowenZhang2020JSAC,CunhuaPan2020TWC} that solve the following problem:
\begin{align}
	\textrm{(P4):} ~~\max_{\mathbf{Q},\mathbf{\Theta} } \quad & \log \det \left( \mathbf{I}_{M_{\sf r}} + \frac{1}{\sigma^2} \mathbf{H}_{\rm eff} \mathbf{Q} \mathbf{H}^H_{\rm eff} \right) \nonumber  \\ 
	\operatorname{ s.t. } \quad
	& \textrm{C12:}~ \theta_{n} \in \left[ 0, 2\pi \right), ~~\forall n \in \mathcal{N}, \nonumber\\
	&\textrm{C1}, \text{C2}. \nonumber
\end{align}
We note that (P4) is a more general form of (P1) by dropping the RIS partitioning constraints.
The PB-element-wise AO method in \cite{ShuowenZhang2020JSAC} iteratively optimizes the transmit covariance matrix $\mbf{Q}$ or one of the reflection coefficients $\theta_n$ ($n=1, \cdots, N$) with the other $N$ blocks fixed.
The PB-WMMSE method in \cite{CunhuaPan2020TWC} solves (P4) in two-layer iterations.
In particular, the outer iteration involves the alternating optimization of $\mbf{Q}$ and $\bsm{\Theta}$, and the inner iteration involves the WMMSE relaxation.
In Table \ref{computational-complexity}, $I_0$ denotes the number of iterations of the PB-element-wise AO method; $I_1$ and $I_2$ denote the numbers of outer and inner iterations of the PB-WMMSE method, respectively.

From Table \ref{computational-complexity}, the computational complexity of the PB-element-wise AO method is cubic with the number of receive antennas $M_{\sf r}$, which is further multiplied by the number of reflecting elements $N$.
The computational complexity of the PB-WMMSE method is both cubic with $M_{\sf t}$ and $N$.
However, the computational complexity of the RIS partitioning based methods is irrelevant to the numbers of transceiver antennas and reflecting elements, and is only polynomial to the number of channel paths for the 1D grid search (best case) and the LM method.
This demonstrates the appealing scalability and low complexity of the proposed algorithms.
	

\section{Simulation Results}\label{sec-simluation}
In this section, we provide numerical results to evaluate the performance of the RIS partitioning based beamforming design.
Consider a simulation scenario where the distances from the Tx to the RIS, from the RIS to the Rx, and from the Tx to the Rx are set as $d_1 = 100$ m, $d_2 = 60$ m, and $d_3 = 150$ m, respectively.
The channel coefficients are generated according to the channel model described in Section \ref{section-system-model}, which is similar to the 3GPP ray-tracing model \cite[Section 7.5]{3GPP}.
In particular, the AoAs/AoDs are uniformly distributed in the continuous angle range, i.e., $(0, \frac{1}{2} \pi ]$ for the elevation angle and $(0, 2\pi]$ for the azimuth angle.
The complex gains $\left\{\alpha_\ell\right\}_{\ell=1}^{L_1}$, $\left\{\beta_\ell\right\}_{\ell=1}^{L_2}$, and $\left\{\gamma_\ell\right\}_{\ell=1}^{L_3}$ are generated from the CSCG distribution with zero mean and unit variance, and then rearranged in the respective descending order in magnitude.
The numbers of dominant channel paths are given by $L_1 = 5$, $L_2 = 7$, and $L_3 = 4$.
Unless otherwise specified, we adopt the default values of the system parameters provided in Table \ref{system-parameters}.
It is worth noting that unlike many existing works assuming that the direct link is completely blocked or very weak \cite{ChongwenHuang2019TWC, QingqingWu2019TWC, ShuowenZhang2020JSAC, CunhuaPan2020TWC, HuayanGuo2020WSR, ccai2021wcl}, we consider a more typical scenario where the path-loss exponents are equally set to $2.4$ for both the cascaded and the direct links.
The experiments are carried out on a Windows x64 machine with 2.90 GHz CPU and 16 GB RAM by MATLAB R2021b.
All results are obtained by averaging over $1000$ independent channel realizations.

\begin{table*}[t]
	\caption{Default Values of Simulation Parameters}
	\label{system-parameters}
	\centering
		\begin{tabular}{||c|c||c|c||}
			\hline
			Parameter & Value & Parameter & Value  \\ \hline
			Number of Tx antennas & $M_{\sf t} = 32$ & Transmission bandwidth & $B = 251.1886$ MHz \\
			\hline
			Number of Rx antennas & $M_{\sf r} = 32$ & Noise power & $\sigma^2 = -90$ dBm \\
			\hline
			Number of reflecting elements & $N = 30 \times 90$ & Transmit power & $P = 30$ dBm \\
			\hline
			Antenna/element spacing & $d = \frac{1}{2} \lambda$ & Path loss for the cascaded channel \cite{WankaiTang2020TWC_PathLoss} & $\mathrm{PL}^{\sf r} =\frac{\lambda^2}{64 \pi^3 d_{1}^{2.4} d_{2}^{2.4} }$ \\
			\hline
			Carrier frequency & $f = 28$ GHz & Path loss for the Tx-Rx channel \cite{WankaiTang2020TWC_PathLoss} & $\mathrm{PL}^{\sf d} =\frac{\lambda^2}{16 \pi^2 d_{3}^{2.4} }$ \\
			\hline
		\end{tabular}
	\end{table*}

\begin{figure*}[t]
	\centering
	\subfigure[]
	{\includegraphics[width=.9\columnwidth]{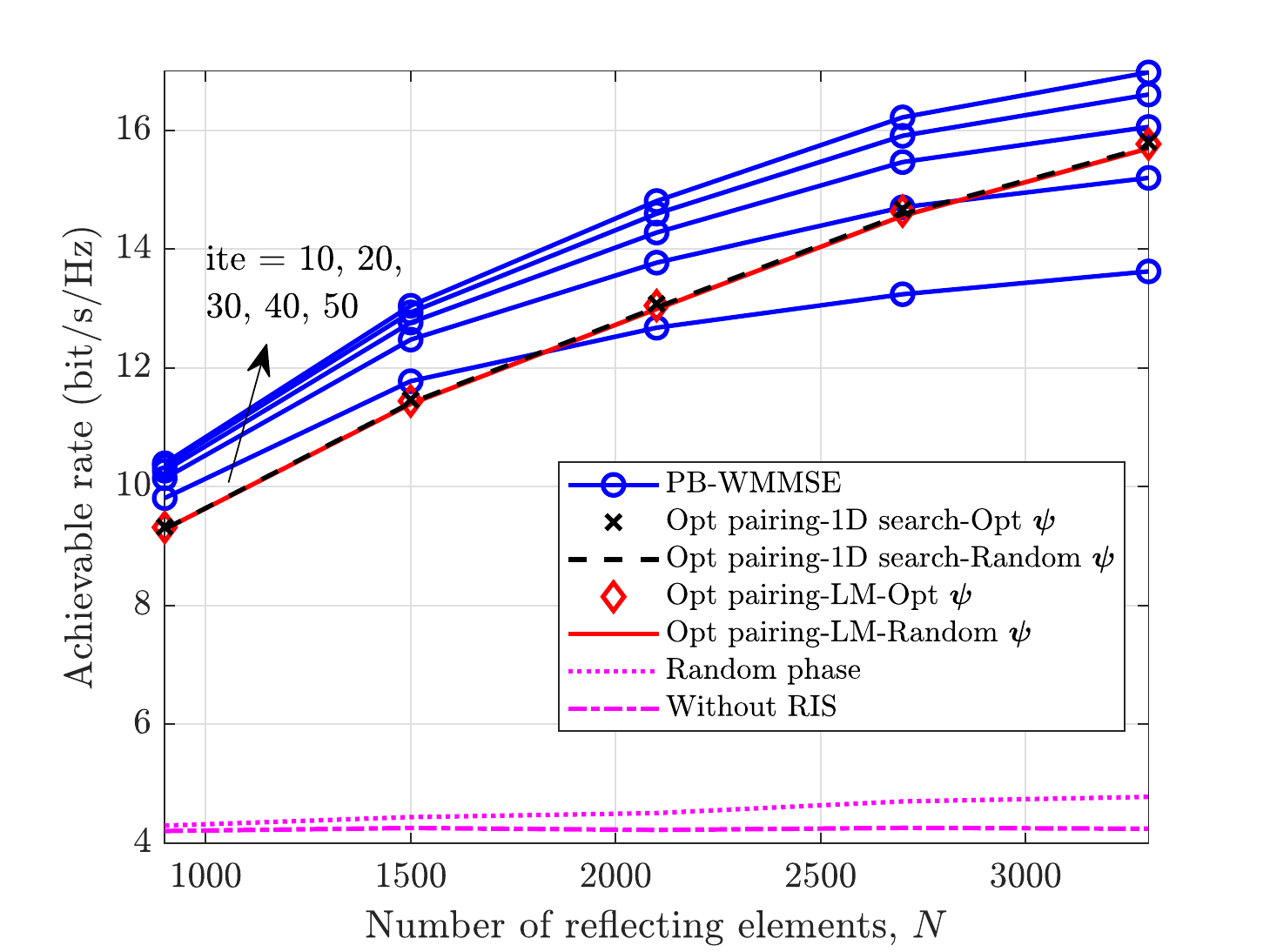}} \hfil
	\subfigure[]
	{\includegraphics[width=.9\columnwidth]{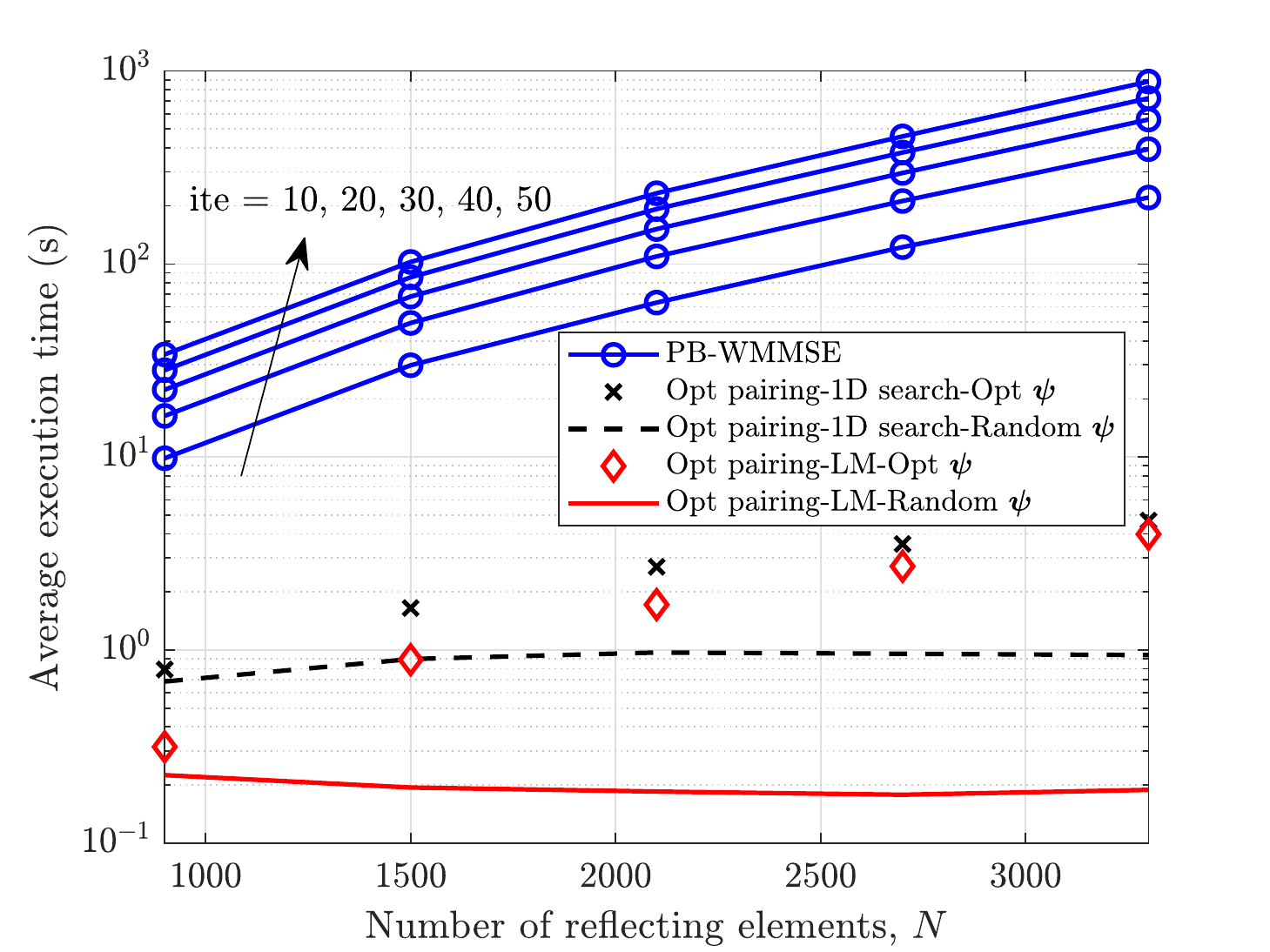}}
	\caption{(a) Achievable rate; (b) average execution time vs. the number of reflecting elements $N$.}
	\label{fig-N}
\end{figure*}

In Fig. \ref{fig-N}, we compare the performance and complexity of the proposed design with the baseline PB-WMMSE method \cite{CunhuaPan2020TWC}.\footnote{The PB-element-wise AO method is not included for comparison in simulations due to its slow convergence rate in large-scale systems.}
Because of the heavy computation burden of the WMMSE method, we terminate the algorithm after $50$ outer iterations, and plot the results when the number of iterations $\mathrm{ite} = 10, 20, 30, 40, 50$.
At each outer iteration, the inner iteration stops when the increase of the target function is less than $10^{-4}$, or the maximum number of inner iterations (set to $1000$) is reached, whichever comes earlier.
For the proposed design, we consider four different implementations of the algorithm, where $\mbf{B}$ is obtained by the optimal path pairing strategy given in Proposition \ref{opt-path-pairing}, $\mbf{p}$ and $\mbf{t}$ are obtained by solving \eqref{nonlinear-equations} either via the 1D search or via the LM method, and $\bsm{\psi}$ is obtained either by the WMMSE method or by random setting.
From Fig. \ref{fig-N}(a), we observe that all the four implementations have comparable performance with the PB-WMMSE benchmark.
For a moderate number of reflecting elements, e.g., $N=900$, $30$ outer iterations are enough for the PB-WMMSE method to converge,
while for $N=3600$, the performance improvement is still visible after $40$ outer iterations.
This implies that the convergence speed of the PB-WMMSE benchmark slows down as $N$ becomes large, which further increases the computation burden.

We plot the average execution times of different algorithms against $N$ in Fig. \ref{fig-N}(b).
The average execution times required by our proposed designs are all less than $5$ seconds even when $N=3600$, while it takes the PB-WMMSE method about $1000$ seconds to conduct $50$ outer iterations.
Such a substantial complexity reduction makes our proposed method a more practical choice for RIS-aided MIMO communications.
Remarkably, the proposed LM implementation with random $\bsm{\psi}$ has an extremely low complexity, namely less than $0.23$ second to finish execution for all plotted $N$, and at the same time achieves almost the same performance with the other three proposed implementations in the large-scale MIMO scenario.

\begin{figure*}[t]
	\centering
	\subfigure[]
	{\includegraphics[width=.9\columnwidth]{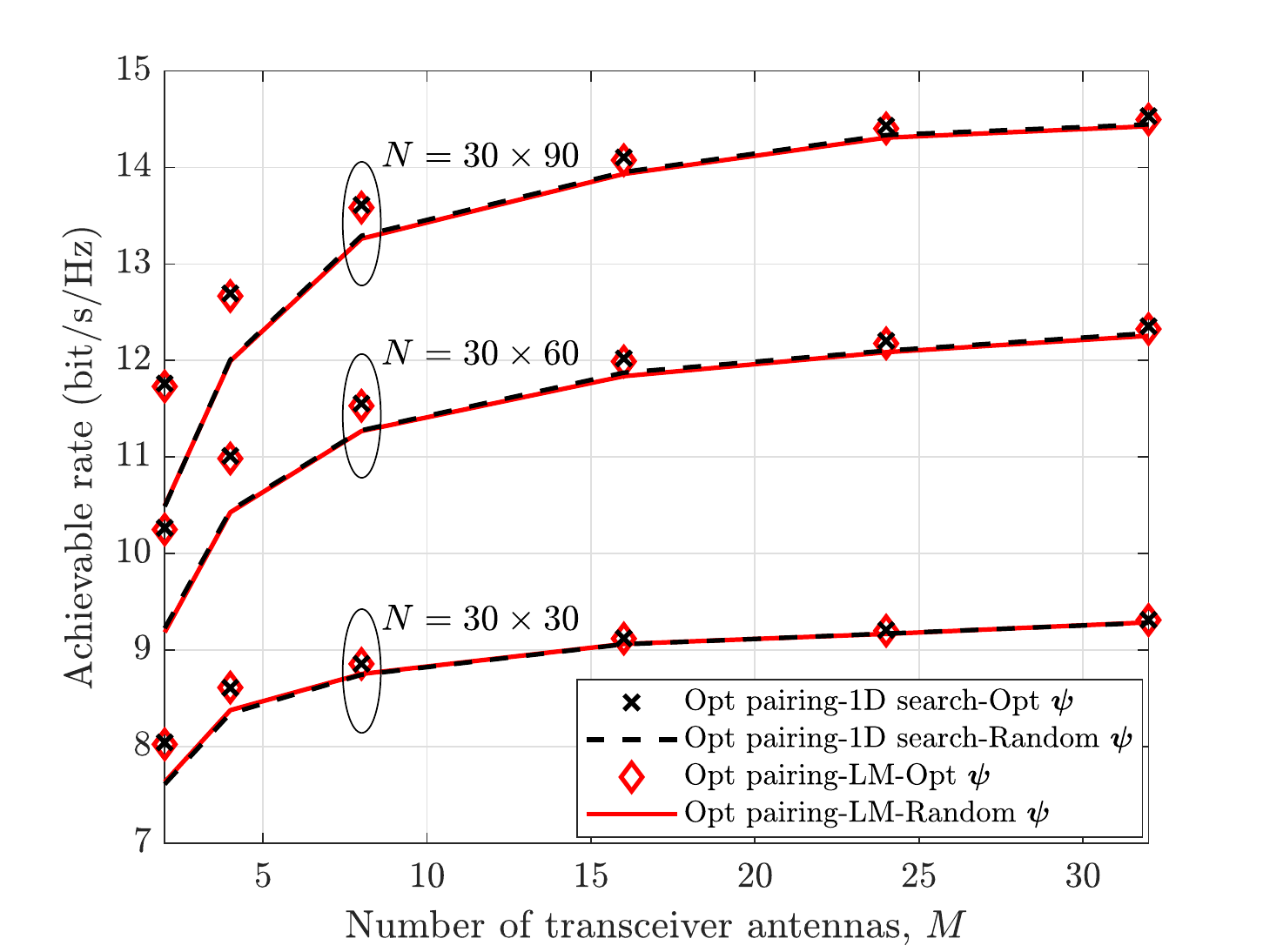}} \hfil
	\subfigure[]
	{\includegraphics[width=.9\columnwidth]{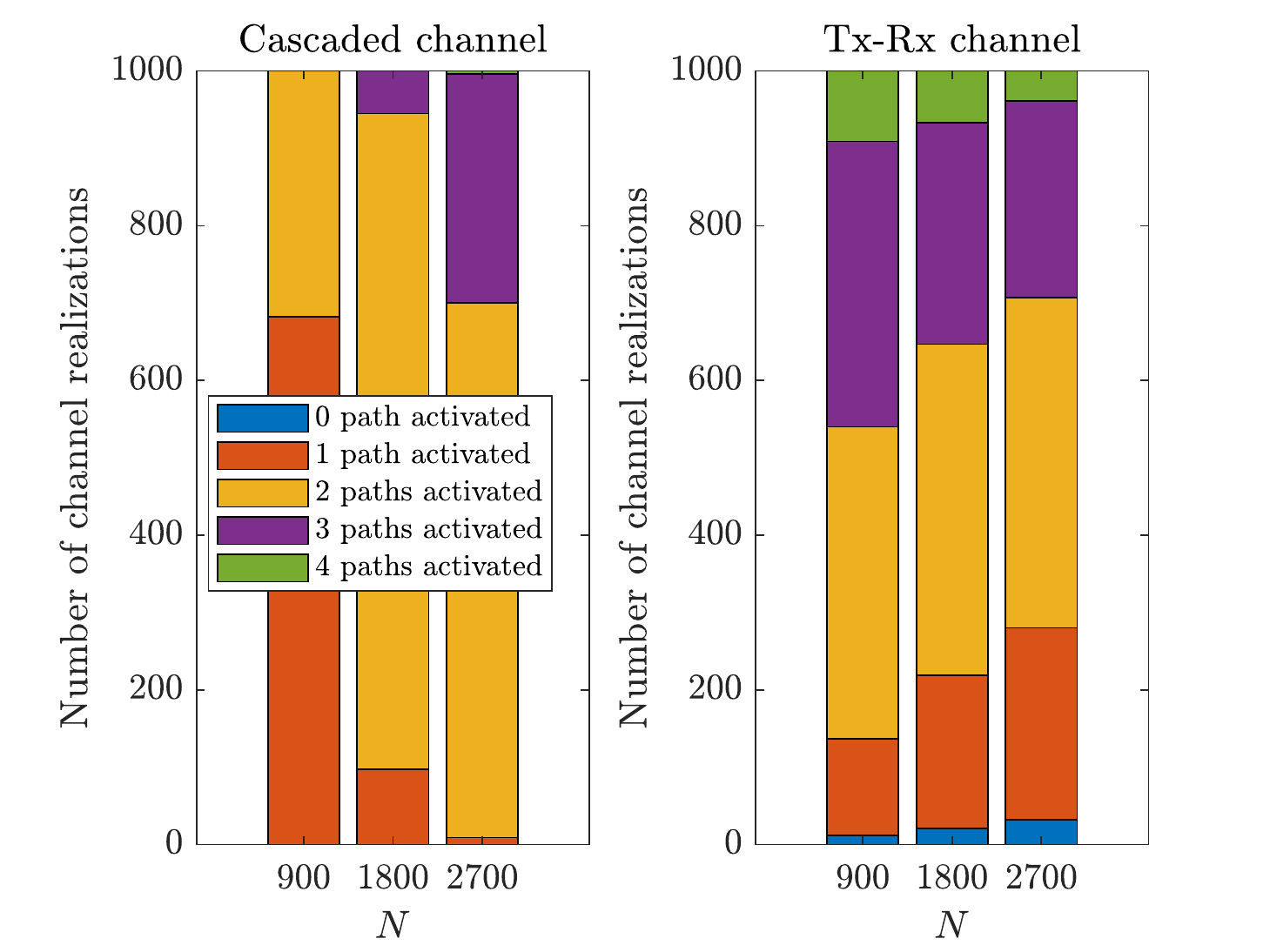}}
	\caption{(a) Achievable rate vs. the number of transceiver antennas $M$, where we set $M_{\sf t} = M_{\sf r} = M$, and the transmit power is set as $P = \frac{P_0}{M_{\sf t} M_{\sf r}}$ with $P_0 = 60.1030$ dBm.
	(b) the occurrence times of different numbers of activated paths in the cascaded channel and the Tx-Rx channel vs. the number of reflecting elements $N$ for the LM method, where $M = 16$.}
	\label{fig-M}
\end{figure*}

\begin{figure*}[t]
	\centering
	\subfigure[]
	{\includegraphics[width=.9\columnwidth]{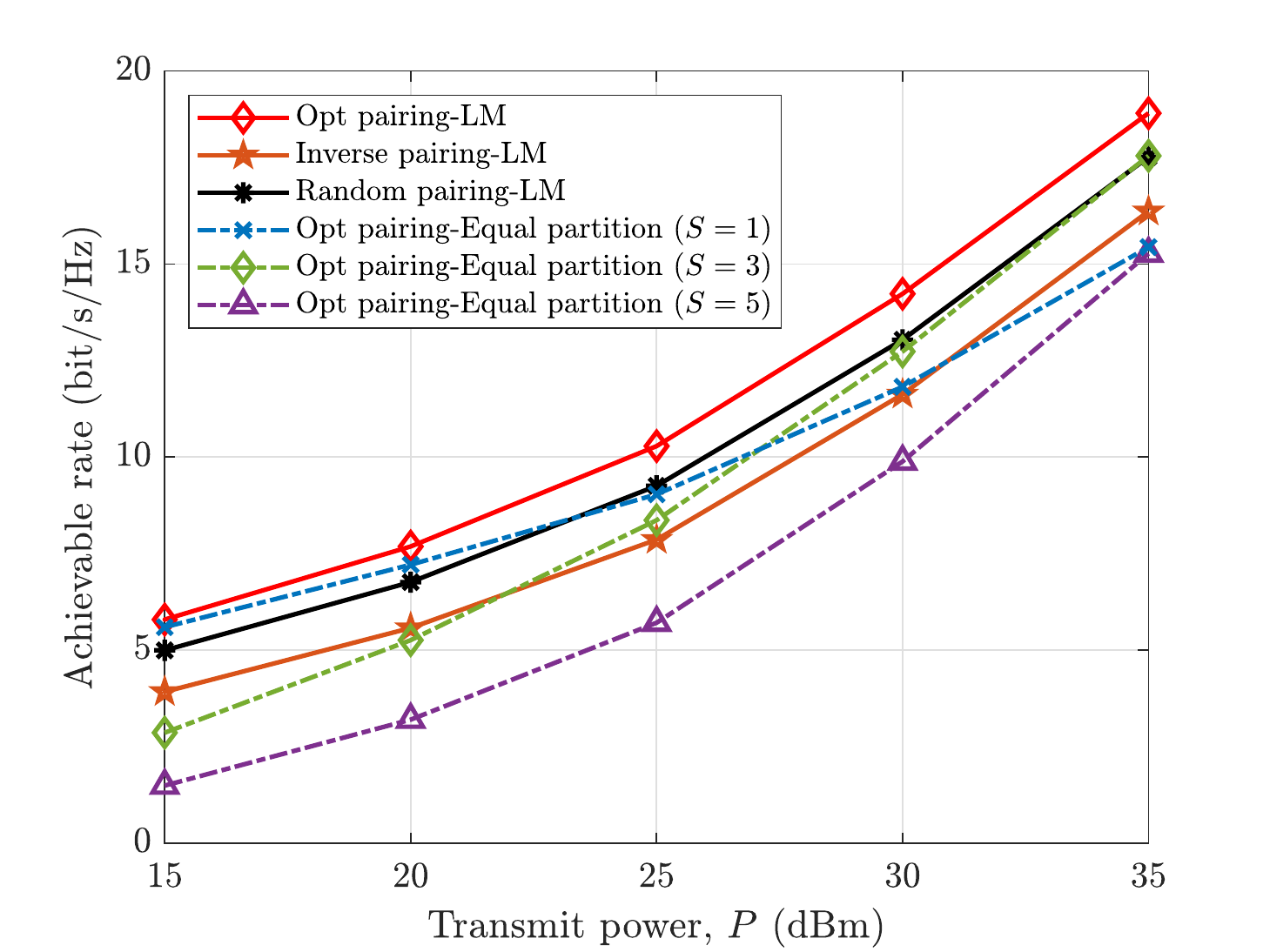}} \hfil
	\subfigure[]
	{\includegraphics[width=.9\columnwidth]{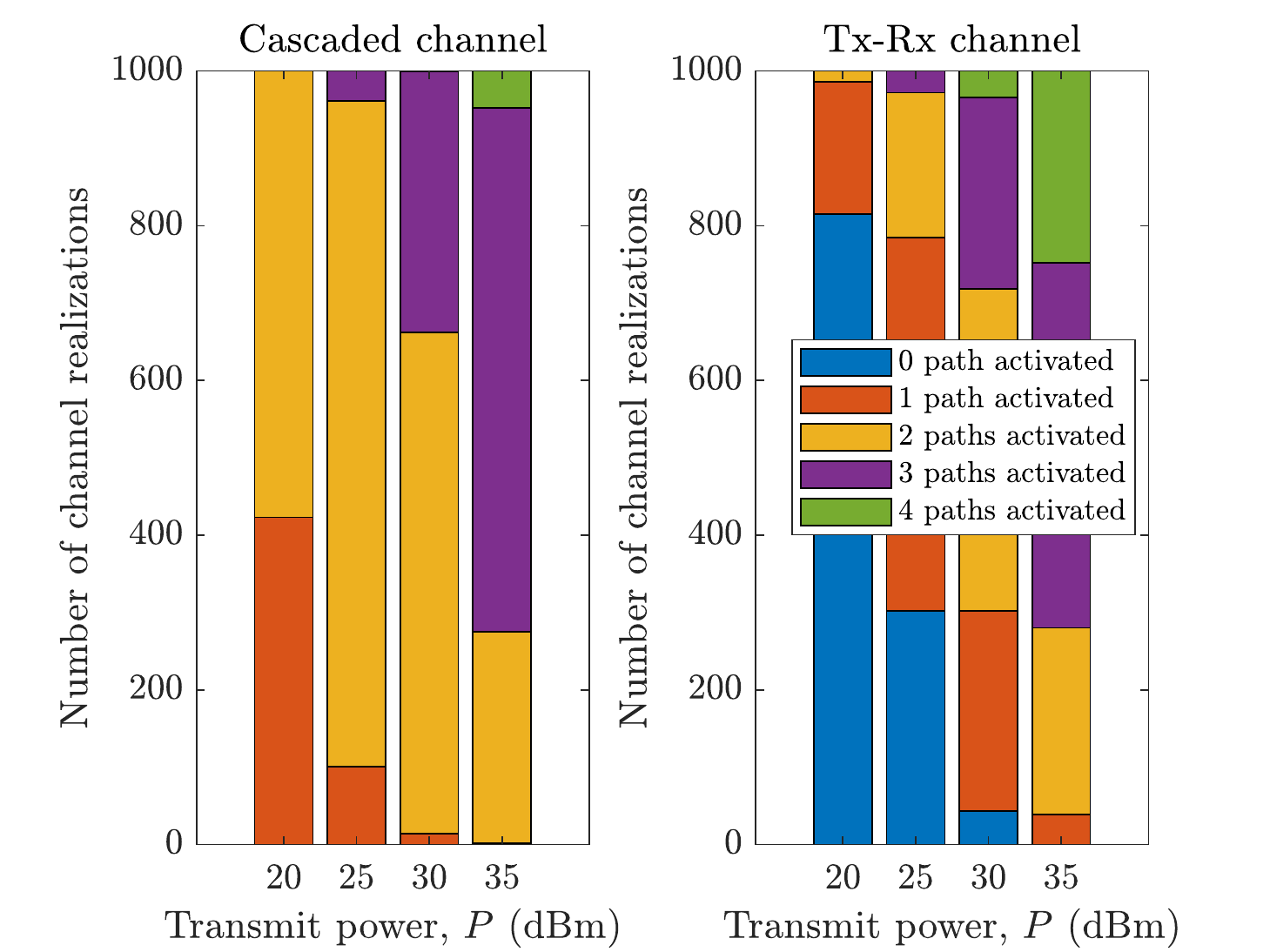}}
	\caption{(a) Achievable rate vs. the transmit power $P$. All the curves are obtained with optimized $\bsm{\psi}$. \color{black}(b) the occurrence times of different numbers of activated paths in the cascaded channel and the Tx-Rx channel vs. the transmit power $P$ for the LM method.}
	\label{fig-P}
\end{figure*}

Fig. \ref{fig-M}(a) plots the achievable rate against the number of transceiver antennas $M$ for different $N$, where we set $M_{\sf t} = M_{\sf r} = M$.
The transmit power is normalized by $M_{\sf t}$ and $M_{\sf r}$ as $P = \frac{P_0}{M_{\sf t} M_{\sf r}}$, where $P_0 = 60.1030$ dBm.
It is observed that solving \eqref{nonlinear-equations} either by the 1D search or by the LM method yields almost the same performance, and the achievable rate differences are mainly brought by different methods for optimizing $\bsm{\psi}$.
In particular, the WMMSE method for optimizing $\bsm{\psi}$ outperforms the random $\bsm{\psi}$ setting when $M$ is relatively small.
The reason is that when $M$ is small, the transceiver antenna arrays do not have enough spatial resolution to distinguish signals from different paths.
As $\bsm{\psi}$ controls the wavefront phase of the signal transmitted through each paired Tx-RIS-Rx path, it is crucial to choose an appropriate $\bsm{\psi}$ to minimize the inter-path interference caused by limited spatial resolution.

To gain more insights, for the case of $M=16$ in Fig. \ref{fig-M}(a) employing the LM method, we provide in Fig. \ref{fig-M}(b) the occurrence times of different numbers of activated Tx-RIS-Rx and Tx-Rx paths in the simulated $1000$ independent channel realizations.
It is shown that as $N$ increases,
the system tends to activate more paths in the cascaded channel for data transmission.
For instance, for $N=900$,
the occurrence times for one, two, three and four activated paths in the cascaded channel are $682$, $318$, $0$, and $0$, respectively;
for $N=2700$, the corresponding occurrence times become $9$, $691$, $296$, and $4$.
In contrast, less Tx-Rx paths tend to be activated as $N$ grows.
This is because a larger $N$ provides a better link quality for the paths in the cascaded channel.
Thus, for a given constant power budget $P$, more power is dedicated to the transmission through the cascaded channel, and hence more paths in the cascaded channel are activated to support multiplexing.

In Fig. \ref{fig-P}(a), we show the achievable rate versus the transmit power $P$ in various path pairing approaches.
It is observed that the proposed optimal path pairing strategy has a superior performance compared to the inverse pairing case in which the Tx-RIS and RIS-Rx paths are paired inversely based on their own path gains, and the random pairing case in which the Tx-RIS and RIS-Rx paths are paired randomly.
This is in agreement with Proposition \ref{opt-path-pairing}.
In addition, we consider a fixed equal partition of RIS, where $S$ equal-sized sub-surface are assigned to serving distinct Tx-RIS-Rx path pairs, chosen from the $S$ largest path pairs in magnitudes.
We observe that there is a noticeable performance gap between the fixed equal partition when $S=5$ and the proposed adaptive resizing by the LM method.
The reason is that the case of $S=5$ allocates much reflecting resource to serve the weak paths.
For the case of $S=1$, the RIS is dedicated to only serving the strongest Tx-RIS and RIS-Rx paths.
We see that the performance gap between the fixed partition of $S=1$ and the LM method is negligible when $P$ is small.
This can be explained in Fig. \ref{fig-P}(b), which depicts the occurrence times of different numbers of activated paths versus the transmit power $P$ employing the LM method.
In the low SNR regime, e.g., $P = 20$ dBm, the LM method only activates one or two paths in the cascaded channel for data transmission, where the fixed partition of $S=1$ is close to optimal.
However, as the transmit power increases, more paths in the cascaded channel need to be activated, which leads to an increased gap between the LM method and the fixed partition of $S=1$.

\section{Conclusions} \label{sec-conclusion}
In this paper, we proposed a RIS-partitioning-based scalable beamforming design for RIS-aided large-scale MIMO systems.
We formulated the achievable rate maximization problem by jointly optimizing active and passive beamforming, where the passive beamforming optimization reduces to the manipulation of the sub-surface sizes, the phase gradients of sub-surfaces, and the common phase shifts of sub-surfaces.
We first focused on the asymptotic regime where the numbers of transceiver antennas and RIS elements go to infinity.
The asymptotic formulation of the problem yields a clear and simple form, which allows to characterize the fundamental performance-complexity tradeoff of RIS partitioning.
Moreover, we characterized the asymptotically optimal solution via a set of non-linear scalar equations.
We also presented the 1D gird search algorithm and the LM method to efficiently solve the equations.
Then, we discussed the insights and impacts of the asymptotically optimal solution on finite-size system design. 
Simulation results demonstrated appealing performance and low complexity of the proposed RIS partitioning design.

\appendices
\section{Generalization of the Asymptotic Analysis to 2D Partitioning of RIS}
	\begin{figure}
	[t]
	\centering
	\includegraphics[width=.9\columnwidth]{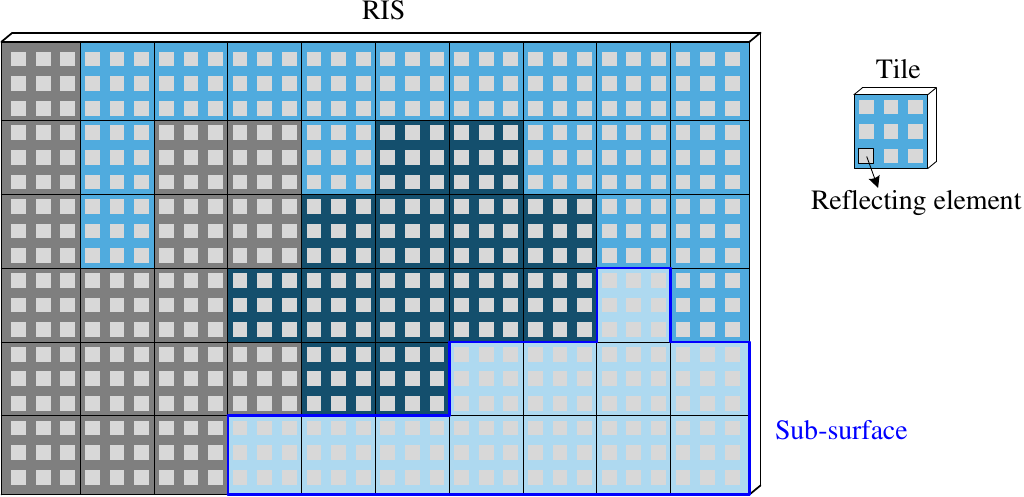}
	\caption{An illustrative example of approximating arbitrary shaped sub-surfaces by multiple tiles.}
	\label{2D_Partitioning_Tile}
\end{figure}

We discuss how to generalize the asymptotic analysis for horizontal partitioning of RIS in Section \ref{sec-asymptotic-formulation} to more arbitrary 2D partitioning of RIS as follows.
As shown in Fig. \ref{2D_Partitioning_Tile}, we partition the RIS into $N^\alpha = N_{\sf x}^{\alpha} \times N_{\sf y}^{\alpha}$ rectangular tiles with $\alpha \in (0,1)$ controlling the tile size.
Each tile consists of $N^{1-\alpha} = N_{\sf x}^{1-\alpha} \times N_{\sf y}^{1-\alpha}$ reflecting elements. 
By merging multiple tiles into a sub-surface to employ the same phase gradient, we obtain more flexible 2D shapes of the sub-surfaces. 
Denote $\mu_s N^{\alpha}$ as the number of tiles owned by the $s$-th sub-surface, where $\mu_s N^{\alpha}$ is assumed to be an integer with $\mu_s \in [0,1]$.
The normalized passive beamforming gain $d_{u,v}$ is expressed as
	\begin{align}
	d_{u,v} &= \frac{1}{N}\sum_{m_{\sf x} = 1}^{N_{\sf x}^{\alpha}} \sum_{m_{\sf y} = 1}^{N_{\sf y}^\alpha} e^{\jmath \psi_{m_{\sf x}, m_{\sf y}}}
	\Bigg(
	\sum_{n_{\sf x}=1}^{N_{\sf x}^{1-\alpha}} e^{\jmath k (n_{\sf x}-1) \eta_{{\sf x},s,u,v}} \nonumber \\
	&~~~~\sum_{n_{\sf y}=1}^{N_{\sf y}^{1-\alpha}} e^{\jmath k (n_{\sf y}-1) \eta_{{\sf y}, s,u,v}}
	\Bigg) \nonumber \\
	&= \frac{1}{N^\alpha} \sum_{m_{\sf x} = 1}^{N_{\sf x}^{\alpha}} \sum_{m_{\sf y} = 1}^{N_{\sf y}^\alpha} e^{\jmath \widetilde{\psi}_{m_{\sf x}, m_{\sf y}}}  \frac{\mathrm{sinc} \left(\frac{k}{2} N_{\sf x}^{1-\alpha} \eta_{{\sf x},s,u,v}\right)}{\mathrm{sinc} \left( \frac{k}{2} \eta_{{\sf x},s,u,v} \right)} \nonumber \\
	&~~~~\frac{\mathrm{sinc} \left(\frac{k}{2} N_{\sf y}^{1-\alpha} \eta_{{\sf y}, s,u,v}\right)}{\mathrm{sinc} \left( \frac{k}{2} \eta_{{\sf y}, s,u,v} \right)} , \label{2D-1}
\end{align}
where $s$ is the sub-surface index of the $(m_{\sf x}, m_{\sf y})$-th tile, $\psi_{m_{\sf x}, m_{\sf y}}$ is the common phase shift of the $(m_{\sf x}, m_{\sf y})$-th tile, and
\begin{align}
	\widetilde{\psi}_{m_{\sf x}, m_{\sf y}} &=  \psi_{m_{\sf x}, m_{\sf y}} + \frac{k}{2} (N_{\sf x}^{1-\alpha}-1) \eta_{{\sf x},s,u,v} \nonumber \\
	&~~~+ \frac{k}{2} (N_{\sf y}^{1-\alpha}-1) \eta_{{\sf y}, s,u,v}. \label{2D-2}
\end{align}
According to \eqref{2D-2}, we properly choose $\psi_{m_{\sf x}, m_{\sf y}}$ to ensure that all the tiles in the $s$-th sub-surface have the same $\widetilde{\psi}_{m_{\sf x}, m_{\sf y}}$, denoted as $\widetilde{\psi}_s$.
Then, \eqref{2D-1} can be recast to the summation of the normalized passive beamforming gains of the sub-surfaces as
\begin{align}
	d_{u,v} = \sum_{s \in \mathcal{S}}  e^{\jmath \widetilde{\psi}_s} \mu_s \frac{\mathrm{sinc} \left(\frac{k}{2} N_{\sf x}^{1-\alpha} \eta_{{\sf x},s,u,v}\right)}{\mathrm{sinc} \left( \frac{k}{2} \eta_{{\sf x},s,u,v} \right)} \frac{\mathrm{sinc} \left(\frac{k}{2} N_{\sf y}^{1-\alpha} \eta_{{\sf y}, s,u,v}\right)}{\mathrm{sinc} \left( \frac{k}{2} \eta_{{\sf y}, s,u,v} \right)}. \label{2D-3}
\end{align}
Eq. \eqref{2D-3} resembles the normalized passive beamforming gain for horizontal partitioning case in \eqref{sum-tile-gain}. 
When $N_{\sf x}, N_{\sf y} \rightarrow \infty$, we have
\begin{align} \label{2D-4}
	\lim_{N_{\sf x} ,N_{\sf y} \rightarrow \infty} d_{u,v} = 
	\sum_{s \in \mathcal{S}}  \mathbbm{1}\{\eta_{{\sf x},s,u,v}, \eta_{{\sf y},s,u,v}\} e^{\jmath \widetilde{\psi}_s} \mu_s.
\end{align}
The only difference between \eqref{2D-4} and \eqref{infty-duv} is that, \eqref{2D-4} replaces $t_s$ in \eqref{infty-duv} by $\mu_s$.
Consequently, the subsequent analysis follows the same steps we elaborated in Section \ref{sec-asymptotic-formulation}.
Moreover, the algorithms developed in Section \ref{sec-algorithm} are amenable to handling the 2D partitioning of RIS as well.

\section{Proof of Lemma \ref{opt-2}}
There are $3^S = 9$ possible patterns of the KKT solutions to (P3.2) for $S=2$.
Firstly, we note that the pattern $\mbf{t}^{\left\{0,0\right\}}$ does not exist since $t_1^0 + t_2^0 = 1$ cannot hold in any circumstance.
We discuss the optimality of the remaining eight patterns subsequently.

\subsection{$\mbf{t}^{\left\{-,0\right\}}$, $\mbf{t}^{\left\{+,0\right\}}$, $\mbf{t}^{\left\{0,-\right\}}$, and $\mbf{t}^{\left\{0,+\right\}}$}
\subsubsection{$\mbf{t}^{\left\{-,0\right\}}$ and $\mbf{t}^{\left\{+,0\right\}}$}
Substituting the expression of $\mbf{t}^{\left\{-,0\right\}}$ into C3 yields
\begin{align} \label{+0}
	1 - \sqrt{1 - \frac{w^2}{\widetilde{m}_1}}  = w ,
\end{align}
where $0 < w \leq \sqrt{\widetilde{m}_1}$.
It can be readily verified that the above equation has a solution when $0< \widetilde{m}_1 \leq 1$,
and the solution is unique due to the monotonicity of \eqref{+0}.
Similarly, the pattern $\mbf{t}^{\left\{+,0\right\}}$ exists when $\widetilde{m}_1 > 1$ and also corresponds to a unique KKT solution.
We remark that $\mbf{t}^{\left\{-,0\right\}}$ and $\mbf{t}^{\left\{+,0\right\}}$ correspond to the same solution $\mbf{t} = [1, 0]^T$ even though they have different patterns.

\subsubsection{$\mbf{t}^{\left\{0,-\right\}}$ and $\mbf{t}^{\left\{0,+\right\}}$}
Similarly, the existence conditions of the patterns $\mbf{t}^{\left\{0,-\right\}}$ and $\mbf{t}^{\left\{0,+\right\}}$ are given as $0< \widetilde{m}_2 \leq 1$ and $\widetilde{m}_2 > 1$, respectively.
Also, they correspond to the same solution $\mbf{t} = [0,1]^T$.
However, since $\widetilde{m}_2 \leq \widetilde{m}_1$, the achievable rate at $\mbf{t} = [0, 1]^T$ cannot be greater than that at $\mbf{t} = [1, 0]^T$, i.e., $\log \left(1+\widetilde{m}_2 \right) \leq \log \left(1+\widetilde{m}_1\right)$.

\subsection{$\mbf{t}^{\left\{-,+\right\}}$, $\mbf{t}^{\left\{+,-\right\}}$, $\mbf{t}^{\left\{+,+\right\}}$, and $\mbf{t}^{\left\{-,-\right\}}$}
\subsubsection{$\mbf{t}^{\left\{-,+\right\}}$}
Substituting the expression of the pattern $\mbf{t}^{\left\{-,+\right\}}$ into C3 yields
\begin{align}\label{-+}
	2 - \sqrt{1 - \frac{w^2}{\widetilde{m}_1}} + \sqrt{1 - \frac{w^2}{\widetilde{m}_2}} = w ,
\end{align}
where $0 < w \leq \sqrt{\widetilde{m}_2}$.
It can be readily verified that the above equation has solution when 
\begin{align} \label{-+ec}
	2-\sqrt{1-\frac{\widetilde{m}_2}{\widetilde{m}_1}} < \sqrt{\widetilde{m}_2},
\end{align}
and the solution is unique due to the monotonicity of \eqref{-+}.
Thus, we use $\mbf{t}^{\left\{-,+\right\}}$ in the following to represent the corresponding KKT solution without causing ambiguity.
Next, we prove that $\mbf{t}^{\left\{-,+\right\}}$ is not a globally optimal solution for any $\widetilde{m}_1$ and $\widetilde{m}_2$, provided that the existence condition in \eqref{-+ec} is satisfied.
In this regard, we treat \eqref{-+} as an implicit function of $w$ w.r.t. $\widetilde{m}_1$ and $\widetilde{m}_2$.
The partial derivative $\partial w /\partial \widetilde{m}_1$ can be expressed as
\begin{align} \label{-+partial}
	\frac{\partial w}{\partial \widetilde{m}_1} 
	&= \frac{w^2}{2 \widetilde{m}_1 \sqrt{\widetilde{m}_1^2 - \widetilde{m}_1w^2} } \nonumber \\
	&~~~\left( \frac{w}{\sqrt{\widetilde{m}_1^2 - \widetilde{m}_1w^2}} - \frac{w}{\sqrt{\widetilde{m}_2^2-\widetilde{m}_2 w^2}} - 1 \right)^{-1}.
\end{align}
It can be observed from \eqref{-+partial} that ${\partial w}/{\partial \widetilde{m}_1}<0$ since $\frac{w}{\sqrt{\widetilde{m}_1^2 - \widetilde{m}_1w^2}} - \frac{w}{\sqrt{\widetilde{m}_2^2-\widetilde{m}_2 w^2}} \leq 0$.
Then, we have
\begin{align}
	\frac{\partial t_1^-}{\partial \widetilde{m}_1} = -\frac{\partial t_2^+}{\partial \widetilde{m}_1} = -\frac{\partial t_2^+}{\partial w} \frac{\partial w}{\partial \widetilde{m}_1} <0.
\end{align}
where the first step is due to $t_1^- + t_2^+ = 1$, and the second step is from the chain rule.
It can be shown that $t_1^- t_2^+$ is monotonically decreasing as the increase of $\widetilde{m}_1$, since
\begin{align} \label{-+partial_result}
	\frac{\partial \left(t_1^- t_2^+\right)}{\partial \widetilde{m}_1} &= t_2^+ \frac{\partial t_1^-}{\partial \widetilde{m}_1} + t_1^-\frac{\partial t_2^+}{\partial \widetilde{m}_1} \nonumber \\
	&= \left( t_2^+ - t_1^- \right) \frac{\partial t_1^-}{\partial \widetilde{m}_1} < 0.
\end{align}
Since $\widetilde{m}_1 \geq \widetilde{m}_2$, 
from \eqref{-+partial_result} we obtain
$t_1^- t_2^+ \leq t_1^- t_2^+ \Big|_{\widetilde{m}_1 = \widetilde{m}_2} = \left( \frac{1}{w} -\sqrt{\frac{1}{w^{2}} - \frac{1}{\widetilde{m}_1}} \right) \left( \frac{1}{w} +\sqrt{\frac{1}{w^{2}} - \frac{1}{\widetilde{m}_2}} \right) = \frac{1}{\widetilde{m}_2} $.
We now show that the achievable rate at $\mbf{t}^{\left\{-,+\right\}}$ cannot be greater than that at $\mbf{t} = [1, 0]^{T}$:
\begin{align}
	C \left( \mbf{t}^{\left\{-,+\right\}} \right) &= \log\left(1+\widetilde{m}_1 (t_1^-)^2\right) + \log\left(1+\widetilde{m}_2 (t_2^+)^2\right) \nonumber \\
	&\leq \log \left( 1+ \widetilde{m}_1 \left((t_1^-)^2 + (t_2^+)^2 + \widetilde{m}_2 (t_1^- t_2^+)^2 \right) \right) \nonumber \\
	&= \log \left( 1+ \widetilde{m}_1 (1-2t_1^-t_2^+ + \widetilde{m}_2 (t_1^- t_2^+)^2 ) \right) \nonumber \\
	&\leq \log \left( 1+ \widetilde{m}_1\right) =C \left( [1, 0]^{T} \right),
\end{align}
where the third step applies the fact that $(t_1^-)^2+(t_2^+)^2 = (t_1^- + t_2^+)^2 - 2t_1^-t_2^+$, and the fourth step is obtained by taking the maximum of the quadratic function $t_1^- t_2^+ \mapsto 1-2t_1^-t_2^+ + \widetilde{m}_2 (t_1^- t_2^+)^2$ over $\big(0, \frac{1}{\widetilde{m}_2}\big]$.
This shows that $\mbf{t}^{\left\{-,+\right\}}$ does not yield a global optimum.

\subsubsection{$\mbf{t}^{\left\{+,-\right\}}$}
When the pattern $\mbf{t}^{\left\{+,-\right\}}$ exists, it corresponds to two KKT solutions.
The achievable rates at the two KKT solutions can be shown to be no greater than that at $\mbf{t} = [1, 0]^{T}$ by following almost the same arguments in discarding $\mbf{t}^{\left\{-,+\right\}}$.
We omit the details here.

\subsubsection{$\mbf{t}^{\left\{+,+\right\}}$}
We prove the pattern $\mbf{t}^{\left\{+,+\right\}}$ corresponds to a local maximum as follows. 
Firstly, the existence condition of $\mbf{t}^{\left\{+,+\right\}}$ is given by
\begin{align} \label{++ec}
	2 + \sqrt{1-\frac{\widetilde{m}_2}{\widetilde{m}_1}} < \sqrt{\widetilde{m}_2}.
\end{align}
It is observed that $\widetilde{m}_2 > 4$ is a necessary condition for the above inequality to hold.
Moreover, if the pattern $\mbf{t}^{\left\{+,+\right\}}$ exists (i.e., the existence condition in \eqref{++ec} is satisfied), the KKT solution employs this pattern is also shown to be unique (for the same reason of the case $\mbf{t}^{\left\{-,+\right\}}$).
Thus, we use $\mbf{t}^{\left\{+,+\right\}}$ in the following to represent the corresponding KKT solution without causing ambiguity.
Substituting $t_2^+ = 1 - t_1^+$ into the objective function of (P3.2) to eliminate $t_2^+$, we recast the problem as 
\begin{align}
	\max_{ 0 \leq t_1^+ \leq 1 } ~   C \left(t_1^+\right) = \log (1+\widetilde{m}_1 (t_1^{+})^2) + \log (1+\widetilde{m}_2 (1-t_1^+)^2).
\end{align}
The second-order derivative of $C\left(t_1^+\right)$ is given by
\begin{align} \label{C-second-order}
	\frac{\partial^2 C}{\partial (t_1^{+})^2} =  \frac{2 \widetilde{m}_1 \left(1- \widetilde{m}_1 (t_1^{+})^2\right)}{\left(1+\widetilde{m}_1 (t_1^{+})^2\right)^2} 
	+  \frac{2 \widetilde{m}_2 \left(1- \widetilde{m}_2 (t_2^{+})^2\right)}{\left(1+\widetilde{m}_2 (t_2^{+})^2\right)^2}.
\end{align}
We observe from $t_2^+ = \frac{1}{w}  + \sqrt{\frac{1}{w^2} - \frac{1}{\widetilde{m}_2}}$ that for any fixed $\widetilde{m}_2$, $t_2^+$ decreases with the increase of $w$.
Based on the monotonicity, we have $t_2^+ \geq {1}/{\sqrt{\widetilde{m}_2}}$ since $w \leq \sqrt{\widetilde{m}_2}$.
This shows that the second term of ${\partial^2 C }/{\partial (t_1^+)^2}$ in \eqref{C-second-order} is non-positive.
Moreover, considering that $t_1^+ = \frac{1}{w}  + \sqrt{\frac{1}{w^2} - \frac{1}{\widetilde{m}_1}}$, $t_2^+ = \frac{1}{w}  + \sqrt{\frac{1}{w^2} - \frac{1}{\widetilde{m}_2}}$, $t_1^+ + t_2^+ = 1$, and $\widetilde{m}_1 \geq \widetilde{m}_2 > 4$, we have $t_2^+ \leq \frac{1}{2} \leq t_1^+$.
Hence, $\widetilde{m}_1 (t_1^+)^2 >  1$ and the first term of ${\partial^2 C }/{\partial (t_1^+)^2}$ is less than zero.
We conclude that ${\partial^2 C }/{\partial (t_1^+)^2} < 0$, which implies that $\mbf{t}^{\left\{+,+\right\}}$ corresponds to a local maximum.

\subsubsection{$\mbf{t}^{\left\{-,-\right\}}$}
The pattern $\mbf{t}^{\left\{-,-\right\}}$ corresponds to a local minimum if it exists.
The proof is similar to the case of $\mbf{t}^{\left\{+,+\right\}}$ and the details are omitted for brevity.

Based on the discussions above, we conclude that when $S = 2$, the optimal solution to (P3.2) occurs only at $\mbf{t}^{\left\{-,0\right\}}$, $\mbf{t}^{\left\{+,0\right\}}$, and $\mbf{t}^{\left\{+,+\right\}}$, in which the first two patterns correspond to the same solution $\mbf{t} = [1, 0]^T$.
This completes the proof of Lemma \ref{opt-2}.

\section{Proof of Proposition \ref{opt-p3b}}
For $S=2$, Proposition \ref{opt-p3b} can be readily proven by Lemma \ref{opt-2}.
Thus, it suffices to consider the case of $S>2$.
We next show that the optimal solution can be obtained by comparing the $S$ solutions given in Proposition \ref{opt-p3b}.
Firstly, if $t_i^0= 0$, we have $t_j^0 = 0$ for all $j>i$ according to Lemma \ref{corol-descend}.
Secondly, we prove by contradiction that the patterns $t_i^+$ and $t_j^-$ ($i \neq j$) cannot co-exist in the optimal solution.
Suppose that the opposite is true.
Then, $[t_i^+, t_j^-]^T$ is necessarily the optimal solution to the following problem:
\begin{subequations} \label{modified-2}
	\begin{align}
		\max_{[t_i, t_j]^T} \quad & \log\left(1+ \widetilde{m}_i t_i^2  \right) + \log\left(1+ \widetilde{m}_j t_j^2  \right) \\ 
		\operatorname{ s.t. } \quad
		& t_i \geq 0,~~ t_j \geq 0, \\
		& t_i + t_j  = 1 - \sum_{k \in \mathcal{S} \backslash \left\{i,j\right\} } t_k.\label{sum-column}
	\end{align}
\end{subequations}
The only difference between the above problem and the problem considered in Lemma \ref{opt-2} lies in the different RIS partitioning budget in \eqref{sum-column}.
The discussion in Lemma \ref{opt-2} can be directly applied here to show that the pattern $[t_i^+, t_j^-]^T$ cannot reach the optimum of the problem in \eqref{modified-2}, which leads to a contradiction.
Thirdly, following similar arguments, we can prove that $t_i^-$ and $t_j^-$ ($i \neq j$) cannot co-exist in the optimal solution.
Based on the above, we conclude that the optimal solution only occurs at $\mbf{t}^{\left\{-, 0,0, \cdots, 0\right\} }$, $\mbf{t}^{\left\{+, 0,0 \cdots, 0\right\} }$, $\mbf{t}^{\left\{+, +, 0,\cdots, 0\right\} }$, $\mbf{t}^{\left\{+, +, +,\cdots, 0\right\} }$, $\cdots$, and $\mbf{t}^{\left\{+, +, +, \cdots, +\right\} }$, where the first two patterns correspond to the same solution $\mbf{t} = [1, 0,0,\cdots, 0]^T$.
Moreover, each pattern corresponds to a unique solution provided that its existence condition is satisfied.
This can be shown by plugging the expressions into C3.
The proof concludes here.

\bibliographystyle{IEEEtran}
\bibliography{IEEEabrv,mybib}

\begin{IEEEbiography}[{\includegraphics[width=1in,height=1.25in,clip,keepaspectratio]{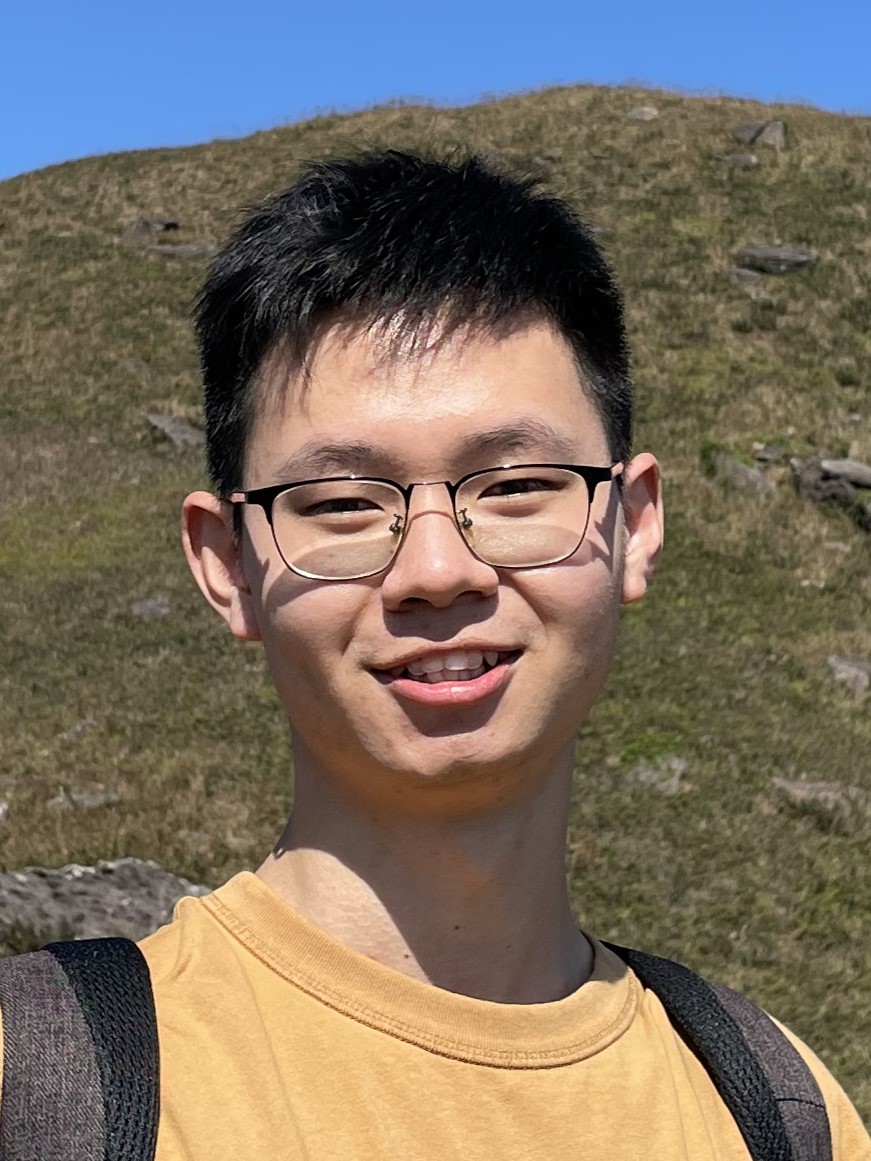}}]{Chang Cai}
	(S’21) is currently a Ph.D. student at the Department of Information Engineering, The Chinese University of Hong Kong (CUHK).
	He received the B.Eng. degree from Yingcai Honors College, University of Electronic Science and Technology of China (UESTC) in 2021.
	
	His research interests lie in the next-generation wireless communication technologies, including but not limited to reconfigurable intelligent surface (RIS) assisted communications, semantic/task-oriented communication, and wireless edge intelligence.
	
	He has been serving as the Managing Editor of IEEE Open Journal of the Communications Society since 2022.
\end{IEEEbiography}

\begin{IEEEbiography}[{\includegraphics[width=1in,height=1.25in,clip,keepaspectratio]{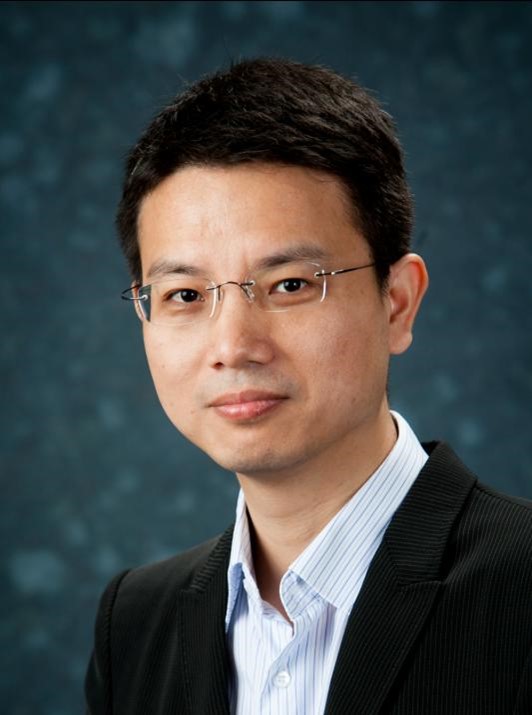}}]{Xiaojun Yuan}
	(S’04-M’09-SM’15) received the Ph.D. degree in Electrical Engineering from the City University of Hong Kong in 2009. From 2009 to 2011, he was a research fellow at the Department of Electronic Engineering, the City University of Hong Kong. He was also a visiting scholar at the Department of Electrical Engineering, the University of Hawaii at Manoa in spring and summer 2009, as well as in the same period of 2010. From 2011 to 2014, he was a research assistant professor with the Institute of Network Coding, The Chinese University of Hong Kong. From 2014 to 2017, he was an assistant professor with the School of Information Science and Technology, ShanghaiTech University. He is now a state-specially-recruited professor with the University of Electronic Science and Technology of China.
	
	His research interests cover a broad range of signal processing, machine learning, and wireless communications, including but not limited to intelligent communications, structured signal reconstruction, Bayesian approximate inference, distributed learning, etc. He has published over 220 peer-reviewed research papers in the leading international journals and conferences in the related areas. He has served on several technical programs for international conferences. He was an editor of IEEE leading journals, including IEEE Transactions on Wireless Communications and IEEE Transactions on Communications. He was a co-recipient of the Best Paper Award of IEEE International Conference on Communications (ICC) 2014, a co-recipient of the Best Journal Paper Award of IEEE Technical Committee on Green Communications and Computing (TCGCC) 2017, and a co-recipient of IEEE Heinrich Hertz Award for Best Communication Letter 2022.
\end{IEEEbiography}

\begin{IEEEbiography}[{\includegraphics[width=1in,height=1.25in,clip,keepaspectratio]{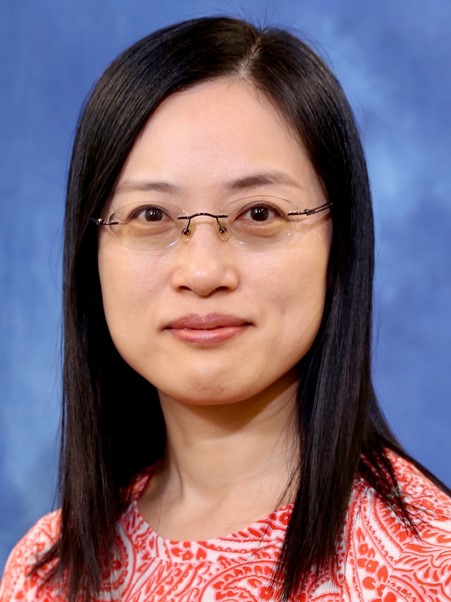}}]{Ying-Jun Angela Zhang}
	(S'00-M'05-SM'10-F'20) received her Ph.D. degree from the Department of Electrical and Electronic Engineering, The Hong Kong University of Science and Technology. She joined the Department of Information Engineering, The Chinese University of Hong Kong in 2005, where is now a professor. Her research interests focus on optimization and learning in wireless communication systems. 

	Prof. Zhang is now a Member-at-Large of IEEE ComSoc Board of Governors, the Editor-in-Chief of IEEE Open Journal of the Communications Society, a member of the Steering Committees of IEEE Transactions on Mobile Computing, IEEE Wireless Communication Letters, and IEEE SmartgridComm Conference. Previously, she served as the Chair of the Executive Editor Committee of IEEE Transactions on Wireless Communications and many years on the editorial boards of IEEE Transactions on Wireless Communications, IEEE Transactions on Communications, IEEE JSAC special issues, IEEE IoT Journal special issues, and IEEE Communications Magazine special issues. Prof. Zhang has served on the Organizing Committees of many top conferences, such as IEEE GLOBECOM, ICC, VTC, SmartgridComm, etc. She was the Founding Chair of IEEE ComSoc Technical Committee of Smart Grid Communications.

	Prof. Zhang is a co-recipient of 2021 and 2014 IEEE ComSoc Asia Pacific Outstanding Paper Awards, 2013 IEEE SmartgridComm Best Paper Award, and 2011 IEEE Marconi Prize Paper Award on Wireless Communications. As the only winner from engineering science, Prof. Zhang won the Hong Kong Young Scientist Award 2006, conferred by the Hong Kong Institute of Science.
\end{IEEEbiography}

\end{document}